\documentclass[aps,prd,onecolumn,superscriptaddress,letterpaper,amsmath,amssymb,nofootinbib]{revtex4}

\pdfoutput=1
\usepackage[normalem]{ulem}

\usepackage[dvips,final]{graphicx}
\usepackage{amssymb}
\usepackage{amsmath}
\usepackage{amsfonts}
\usepackage{epsfig}
\usepackage{bm} 
\usepackage[dvipsnames,usenames]{color}
\usepackage{comment}
\usepackage{float}
\usepackage[utf8]{inputenc}
\usepackage[colorlinks=true,linkcolor=blue,urlcolor=blue,citecolor=blue]{hyperref}

\begin{document}

\title{Deep Neural Networks as the Semi-classical Limit of Quantum Neural Networks}

\author{Antonino Marcian\`o}
\email{marciano@fudan.edu.cn}
\affiliation{Center for Field Theory and Particle Physics \& Department of Physics, Fudan University, 200433 Shanghai, China}
\affiliation{Laboratori Nazionali di Frascati INFN, Frascati (Rome), Italy, EU}

\author{Deen Chen} 
\affiliation{College of Design and Innovation, Tongji University, Shanghai, China}

\author{Filippo Fabrocini$^{(*)}$ \footnote{(*) Corresponding authors}}
\email{fabrocini@tongji.edu.cn, corresponding author (*)}
\affiliation{College of Design and Innovation, Tongji University, Shanghai, China}

\author{Chris Fields}
\email{fieldsres@gmail.com}
\affiliation{Caunes Minervois, France}

\author{Enrico Greco$^{(*)}$}
\email{enrico.greco@live.it, corresponding author (*)}
\affiliation{Department of Physics, Fudan University, 200433 Shanghai, China}

\author{Niels Gresnigt}
\email{niels.gresnigt@xjtlu.edu.cn}
\affiliation{Department of Physics, Xi'an Jiaotong-Liverpool University, Suzhou, Jiangsu 215123, China}

\author{Krid Jinklub}
\email{jinklub_krid@tongji.edu.cn}
\affiliation{College of Design and Innovation, Tongji University, Shanghai, China}

\author{Matteo Lulli}
\email{lulli@sustech.edu.cn}
\affiliation{Department of Mechanics and Aerospace Engineering, Southern University of Science and Technology, Shenzhen, Guangdong 518055, China}

\author{Kostas Terzidis}
\email{kostas@tongji.edu.cn}
\affiliation{College of Design and Innovation, Tongji University, Shanghai, China}

\author{Emanuele Zappala}
\email{zae@usf.edu, emanuele.amedeo.zappala@ut.ee}
\affiliation{Institute of Mathematics and Statistics, University of Tartu, Tartu, Estonia, EU}


\begin{abstract}
Our work intends to show that: (1) Quantum Neural Networks (QNN) can be mapped onto spin-networks, with the consequence that the level of analysis of their operation can be carried out on the side of Topological Quantum Field Theories (TQFT); (2) Deep Neural Networks (DNN) are a subcase of QNN, in the sense that they emerge as the semiclassical limit of QNN; (3) A number of Machine Learning (ML) key-concepts can be rephrased by using the terminology of TQFT. Our framework provides as well a working hypothesis for understanding the generalization behavior of DNN, relating it to the topological features of the graphs structures involved.
\end{abstract}

\maketitle


\section{Introduction}
\label{1}
A paradoxical result in \cite{zhang} according to which DNN memorize the training samples by brute force leaves unexplained where the generalization capabilities of DNN come from. This “apparent” paradox, as it has been dubbed in \cite{26}, has led to active discussions by many scholars; see for example \cite{34,19,20,21,22,23,24,30,25,29,8,32,33}. In any case, in our vision, the overall discussion has empirically proved how far the ML community is from building a principled model of DNN and, therefore, understanding their generalization capabilities. 

Quantum machine learning (QML) and quantum algorithms have been employed successfully to obtain significant computational speedup of classical artificial intelligence methods \cite{lovett,tiersch,Carleo}. The opposite approach, i.e. that of applying classical ML techniques to deduce improved quantum algorithms, is also frequently used, e.g. \cite{aimeur}. Quantum Computing (QC) has provided a very deep theoretical background to apply quantum algorithms to quantum computers, and quantum approaches to quantum tasks have recently found profound applications \cite{paparo,schuld,wiebe}. In the present article we are interested in developing a new theoretical background for ML that is based on mathematical notions derived from quantum topology, and traditionally applied in theoretical physics. Specifically, we aim at using Topological Quantum Field Theory (TQFT) to construct a topological notion of neural network, a Topological Quantum Neural Network (TQNN), whose corresponding quantum algorithms provide an algebra/geometric background to explain the issue of generalization in DNNs. We emphasize that such TQNN are more general than QNN models employing fixed arrays of quantum gates, as in e.g. \cite{5, beer}. Our TQNN structure, in practice, possibly provides a computational advantage as a consequence of the fact that the projectors used in \cite{Noui-Perez} naturally implement arbitrarily deep topological neural networks. We will also show that the semi-classical limit of the objects hereby considered can be interpreted as classical DNNs. 

This pathway has been suggested by the analogy with physics. An experiment at the base of the quantum revolution around the beginnings of the 20th century pointed out the existence of the photoelectric effect. As it is notorious, the effect has been explained by Albert Einstein resorting to a corpuscular description of the electromagnetic field, namely to the concept of photons as carriers of “quanta” of light. But, actually, the interpretation of this very seminal experiment clashed with a common perspective on quantum physics, widely spread nowadays even in the physics community, and that relies on the naive assumption that quantum means microscopic and classical macroscopic. 
A rather different pathway consists in moving from a quantum theory, with a tested semi-classical limit that corresponds to the classical theory, and investigating the varieties of predictions that can then falsify the quantum theory. This approach allows new predictive power and more robust experimental corroboration, and it is the approach we will be following within this paper. 

\section{Motivations and theoretical background}
\label{2}

The main problem addressed in this article is that, despite the excellent performance in many different domains, the source of the success of DNN and the reason for their being powerful ML models remain elusive. DNN are still analytically opaque in the sense that they miss a principled model of their operation. This issue has a theoretical relevance and, at the same time, it is extremely urgent from an applicative point of view as well. Indeed, if we wish to trust any application making use of Deep Learning technology, we need to open the “black box” of these architectures. In this sense, a solution to a problem of this kind is also going to have a social impact to the extent it will improve the trustworthiness of AI systems. It has been empirically shown \cite{zhang} that successful DNN can achieve zero training error or very small error when trained on a completely random labeling of the true data. On the other side, the test error is not better than random chance insofar as there is no correlation between the training labels and the test labels. However, as the authors of the paper underline, in this case learning should have been impossible to the extent that the semantics of the training samples has been completely corrupted by the randomization of the labels, with the consequence that training should not converge or slow down substantially. Surprisingly, the training process was largely unaffected by the transformation of the labels. This result seems to leave unexplained the generalization capabilities of DNN. How to explain that DNN are actually able to achieve more than good generalization performances, even though the results of learning a function that maps an input to an output based on example input-output pairs show that the training set has been memorized by brute force?

Moreover, the results of \cite{zhang} have posed a challenge to Computational Learning Theory (CoLT) as well. The experimental results emphasize that the effective capacity of several successful DNN is large enough to shatter the training data. In other words, the capacity of these models is in principle rich enough to memorize the training data (with or without the use of regularizers). In particular, the classical measures of ML model expressivity (VC-dimension, Rademacher complexity, etc.) seem to fail when explaining the capabilities of DNN. Specifically, they do not explain the good generalization behavior achieved by DNN, which are typically over-parametrized models that often have substantially less training data than model parameters \cite{28}. As a matter of fact, it is usually understood that good generalization is obtained when a ML model does not memorize the training data, but rather learns some underlying rule associated with the data generation process, therefore being able to extrapolate that rule from the training data to new unseen data. Overfitting and, even more, brute force memorization should exclude generalization by definition, even as concerns human beings. For instance, the concepts of capacity (\cite{miller,wattenmaker,lewis,cowan2001magical,feldman2000minimization,zhu2009human}),  bias (\cite{griffiths2008using,griffiths2010bayesian}), overfitting (\cite{o1994hippocampal,vong2016additional}), and generalization (\cite{shepard1987toward,kemp2014taxonomy}) have been widely explored in cognitive psychology as well.

This scenario has prompted us towards considering a different framework, the TQNN framework, for revising a number of traditional ML concepts in the light of concepts coming from TQFT. 

We start by considering QNN and pointing out certain fundamental perspectives that will also appear in our topological TQNN, when considering the semi-classical limit. 
A recurrent visual image for QNN involves nodes of the hidden layers that interconnect from each neighbor to another. In our setting, we will not consider fixed topologies of this type, but rather consider $2$-complexes bounding graphs, which are associated to input and output states. 

\begin{figure}
\centering
\includegraphics[width=10 cm]{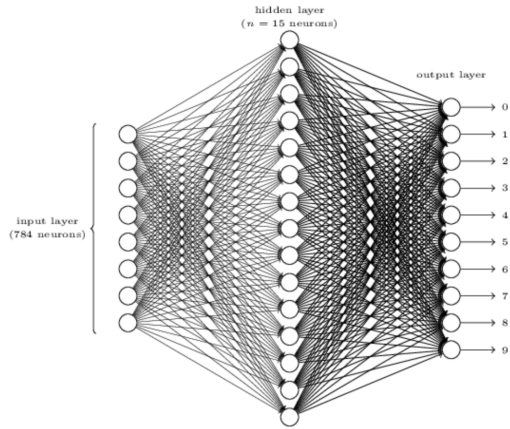}
\caption{A three-layer neural network for recognizing digits.}
\label{Fig:QNN}
\end{figure}

As a starting point to move from, we consider a traditional feedforward architecture (Figure~\ref{Fig:QNN}), as it could be used in classifying individual hand written digits. It is inessential to the goal of this paper to define whether an architecture of this kind will make use of backpropagation or whatever other optimization technique. We assign a set of squared $(2j+1)\times (2j+1)$ matrices, the dimension of which is specified by half-integer numbers $j$, and which depend on the three Euler angles $\phi$, $\theta$ and $\psi$, to the links and to the nodes of a graph. 
The assignment of $(2j+1)\times (2j+1)$ matrices to the links of the DNN is the first step required to introduce the concept of TQNN we are proposing. In the next sections we will consider wide generalizations of this construction in terms of specific mathematical structures that are well known in theoretical physics, namely TQFT.   

The ML task will consist in classifying individual handwritten digits. Figure 1 illustrates the three-layer neural network we could use for recognizing the individual digits. The input layer of the network will contain neurons encoding the values of the input pixels. Our training data will consist of a sample of $28 \times 28$ pixel images of scanned handwritten digits. Therefore, the input layer will contain $784 = 28 \times 28 $ neurons. The input pixels are greyscale, with a value of $0.0$ representing white, a value of $1.0$ representing black, and the values in between representing gradually darkening shades of grey. The second layer of the network is a hidden layer. The example illustrates a hidden layer containing just $n = 15$ neurons. Finally, the output layer of the network will contain $10$ neurons. We will number the output neurons from $0$ through $9$, and figure out which neuron has the highest activation value. If, say, the first neuron will have an output $\approx$1, then that will indicate that the network has classified the digit as $0$.\\

The TQNN associated to the architecture described may be recovered by: i) Selecting, in the bulk of the DNN, a graph with three-valent and four-valent vertices; ii) Associating to the edges interconnecting vertices $(2j+1)\times (2j+1)$ matrices labelled by either $j=1/2$ or $j=1$; iii) Given any three-valent vertex, to the incoming or outgoing three edges of which are assigned matrices (one with dimension labelled by $j_3=1$ and two with dimension specified by $j_1=j_2=1/2$), assigning to it a three-valent tensor saturating the indices of the matrices with the $j_1=j_2=1/2$ and $j_3=1$ (Pauli matrices); iv) Finally, assigning to any vertex in which four $1/2$-colored edges are incoming or outgoing (three edges laying on the same layer and a fourth one external to it) a four-valent intertwiner tensor among the $1/2$-colored matrices (contractions of two Pauli matrices).  The next section discusses this construction in detail.

\section{Topological Quantum Neural Networks}\label{sec:TQNN}

\begin{figure}
\centering
\includegraphics[width=10 cm]{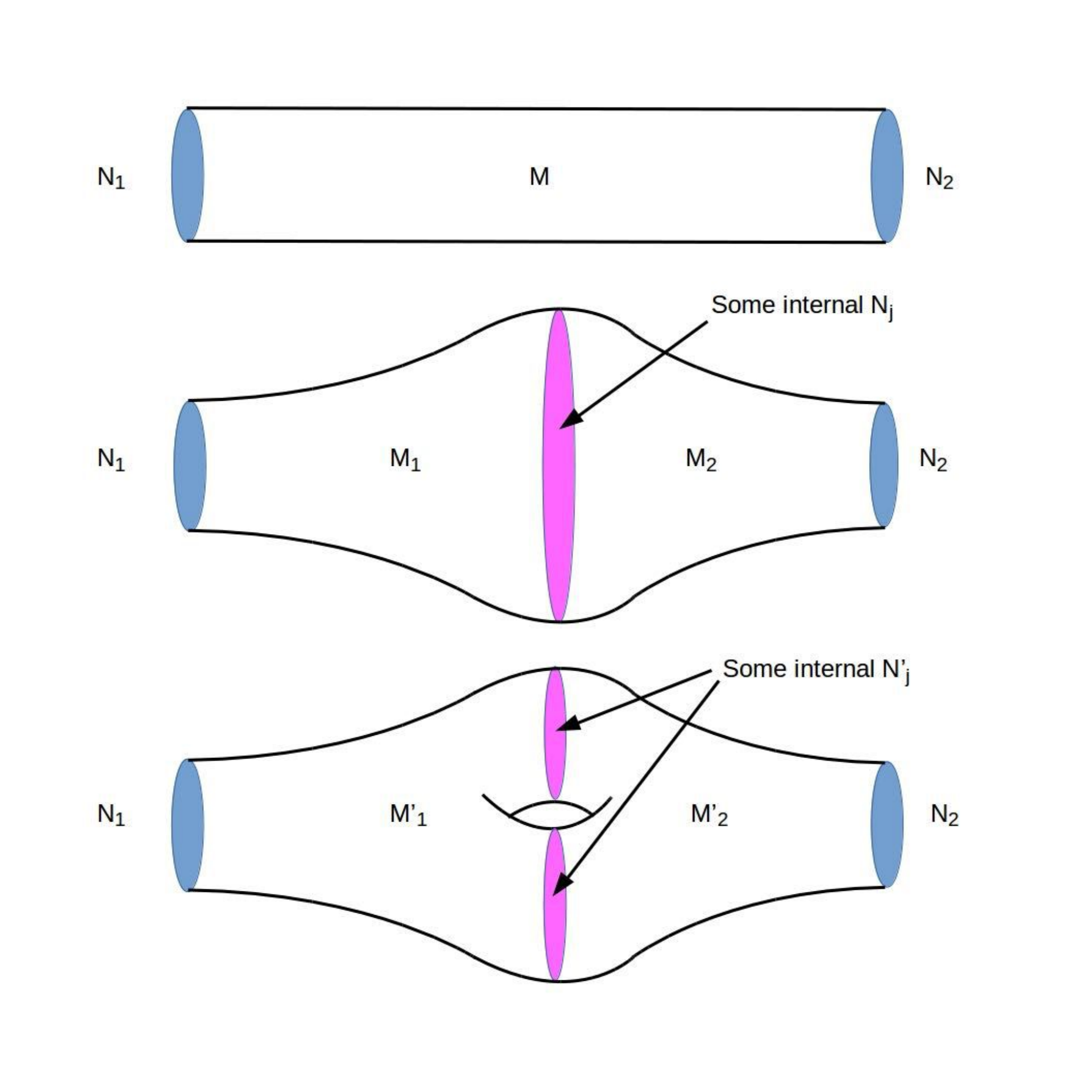}
\caption{Schematic representation of $\mathcal Cob$. The top drawing shows a manifold $M$ whose boundary consists of two manifolds $N_1$ and $N_2$. While $N_1$ and $N_2$ are objects in $\mathcal Cob$, the manifold $M$ is a morphism. In the middle and bottom, two cobordisms are glued along their common boundaries (where the orientation of $N_j$ in $M_2$ is taken with opposite sign). This provides with a composition rule for morphisms having same target and source objects.}
\label{Cob}
\end{figure}

The mathematical structure used to define TQNN is that of TQFT.  Formally, a TQFT is a functor from the category of cobordisms, which we denote by $\mathcal Cob$, to the category of vector spaces. See Figure~\ref{Cob} for a concise description of cobordisms. Roughly speaking, what the definition of TQFT means, is that to each closed $(n-1)$-manifold we associate a vector space (of arbitrary dimension) on some fixed base field, usually $\mathbb C$, and to each $n$-manifold $M$ between two $(n-1)$-manifolds $N_1$ and $N_2$, we associate a linear map between the vector spaces corresponding to $N_1$ and $N_2$. What functoriality encodes in this context is the coherence of composition of manifolds (i.e. gluing manifolds along their boundaries) with respect to composition of linear maps. With respect to Figure~\ref{Cob}, the manifolds $N_1$ and $N_2$ in the top drawing of the figure are associated by a TQFT to vector spaces $V_1$ and $V_2$, while $M$ becomes a linear map between $V_1$ and $V_2$. In the two drawings in the middle and bottom of Figure~\ref{Cob}, the linear maps corresponding to $M_1$ and $M_2$ are composed, through the vector space associated to $N_j$, which we call $V_j$. In the case of the bottom drawing, further, $V_j$ is the tensor product of two vector spaces, corresponding to the two connected components of $N_j$. By functoriality, we have that if $f_i$ and $g_i$ are the maps associated to $M_i$ and $M'_i$, respectively, and $f$ and $g$ are the maps associated to $M_1 \bigcup M_2$ and $M'_1\bigcup M'_2$, respectively, then it holds that $f_2\circ f_1 = f$ and $g_2\circ g_1 = g$. In other words, the composition rule of $\mathcal Cob$ is translated into the composition of linear maps between vector spaces. We can, in particular, think of any linear map $f: N_1 \mapsto N_2$ as an arbitrary finite composition $f = f_m \circ f_{m-1} \circ \dots f_2 \circ f_1$, where each of these $m$ maps is associated to some $n$-manifold $M_k$, subject only to constraint that the $M_k$ can be successively glued together.  Hence we can equally well think of each of the $f_k$ as an equivalence class of smooth paths through $M_k$, paths to which amplitudes will be assigned in the construction below.

The typical elementary example of TQFT is in dimension $2$, i.e. one dimension lower than the TQFTs considered in this article. We have a fixed vector space $V$ for each copy of the circle (i.e. $1$-manifolds), and the vector space $V^{\otimes r}$ is associated to $1$-manifolds that consist of multiple copies of circles. Then, let $N_1$ consist of $r$ circles and $N_2$ of $s$ circles. To a surface connecting $N_1$ and $N_2$ we associate a linear map $V^{\otimes r} \longrightarrow V^{\otimes s}$. 
It is a ``folklore'' result in quantum topology that TQFTs in dimension $2$ are classified by Frobenius algebras. Observe, in particular, that in the previous scheme we have that to a closed manifold (i.e. without boundaries $N_1$ and $N_2$) is associated a linear map between two copies of $V^{\otimes 0} \cong \mathbb C$. This is nothing but a complex number which is an invariant of the manifold. 

The class of TQFT relevant to this article come from quantum gravity, in the holonomy representation, where we have that the boundary vector spaces are Hilbert spaces whose bases are given by cylindrical functions corresponding to spin-networks. We define a TQNN to be a TQFT whose target vector spaces are tensor products of the Hilbert space of cylindrical functions, taken with the (regularized) Ashtekar-Lewandowski metric.

In this setting, therefore, we can take an input spin-network associated to the dual cubulation of a boundary manifold, and map this to another output spin-network. Associated to such a mapping there arises a scalar in the ground field that is geometrically derived by ``capping'' the boundary components to obtain a closed manifold. This scalar is interpreted as being a probability amplitude for a transition between two spin-networks. This is the outcome of applying a TQNN between input and output states. In concrete, a TQNN returns, given two spin-networks $(\Gamma_{in},\Gamma_{out})$, the transition amplitude from $\Gamma_{in}$ to $\Gamma_{out}$, which in turn can be used for a binary classification problem, e.g. a transition amplitude whose modulus square is higher than a predefined ``confidence'' number between $0$ and $1$ implies that the input is classified as the output. 

\begin{figure}
\centering
\includegraphics[width=12 cm]{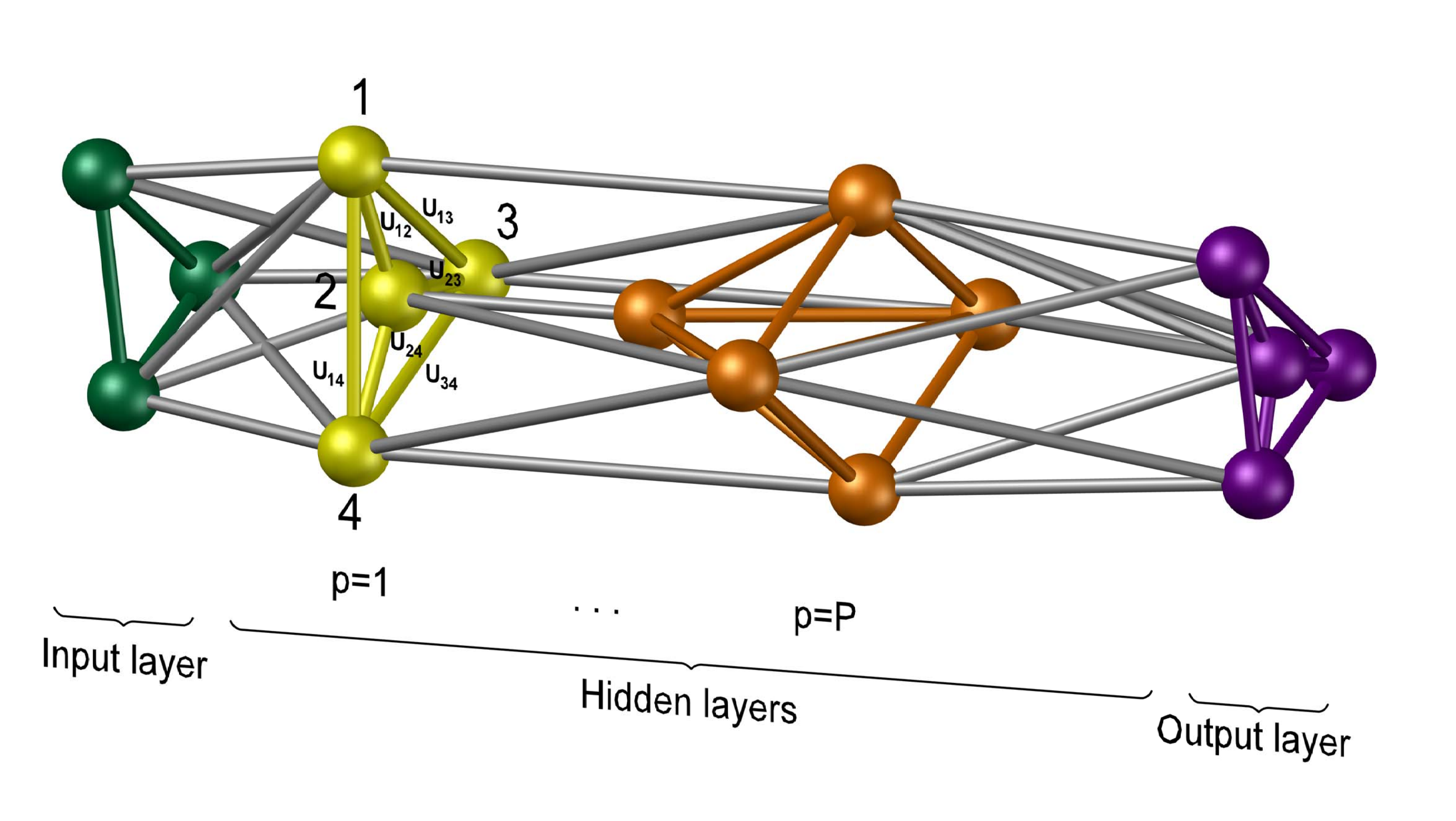}
\caption{A functorial evolution among two spin-network states.}
\label{HL}
\end{figure}

A tight texture of analogies fetched by the equivalence between this categorical approach to quantum field theory and deep machine learning specifies the theoretical perspective through which we progress. Both the Hilbert space states and the probability amplitudes describing their relative transitions are crucial to the individuation of a TQNN capable to include DNN as a specific sub-case. Following the recent literature \cite{5}, these states can be considered as QNN machines, and their state transitions as implementing quantum computations. These are supported on 1-complexes (graph $\Gamma$), and are endowed with a functorial evolution supported on 2-complexes. This 2-complex evolution is in turn a cobordism acting at an internal boundary (an $n-1$-manifold) that is effectively a ``hidden layer'' of the TQNN; however unlike in a QNN architecture with fixed layers, in a TQNN each ``layer'' can be further decomposed into a (finite) sequence of intermediate evolution operators ($n$-manifolds glued by further cobordisms) and hence into a further nested sequence of ``hidden layers'' as schematized in Figure~\ref{HL}. As we will see, this functorial evolution on 2-complexes is amendable to a training algorithm specifically adapted to our TQNN framework.

 We consider, in the present article, the case of TQFT with a local non-abelian Lie group, which we assume for the sake of simplicity to be SU$(2)$. This specific choice, in particular, allows us to parallel the example of QNN provided above. Then, squared $(2j+1)\times (2j+1)$ matrices depending on the Euler angles turn out to constitute the representations of the group elements $U\in$ SU$(2)$. Tensors saturating, at the vertices, the matrix indices are here specified by the intertwiners of SU$(2)$. 
 In our setting, these are initial and final states of the TQNN, rather than the network itself. 
 The functor, as an operator the action of which is supported on the disjoint boundary states, corresponds to the classifier, i.e. the overall map $f: N_1 \mapsto N_2$ implemented by the TQNN as described above. The scheme of computing the transition amplitude between initial and final states is obtained following an association path \cite{Rovelli2010}. This is schematically described as follows.

\begin{itemize}
\item
We integrate either twice over each internal edge in the bulk \footnote{For bulk we intend any 2-complex structure, without boundary. Therefore $Z_\mathcal{C}[U_{\gamma_l}]$ acts in a functorial way on a the boundary states, which are 1-complexes, i.e. colored graphs $\Gamma$ composed by a collection of paths $\gamma$ and nodes where the paths intersect, to which are assigned respectively holonomies and intertwiners.}, or once over adjacent couple of group elements, assigned to either internal edges or vertices:  
\begin{equation}
\begin{picture}(35,25)
\thicklines
\put(-7,-11){\tiny $U'$}
\put(22,16){\tiny $U$}
\put(0,-6){\line(1,1){20}}
\put(10,-1){\tiny $e$}
\end{picture} \ \ 
\qquad  \qquad    \Longrightarrow
\qquad  \qquad  
\int_{\rm SU(2)} d U_{s_e}\int_{\rm SU(2)}d U_{t_e}\,;
\label{uno}
\end{equation}

\item 
We integrate over each couple of adjacent group element, assigned to either to a face or to an internal edge:
\begin{equation}
\begin{picture}(35,25)
\put(0,-6.2){\line(-1,0){10}}
\put(20.2,14){\line(-1,1){10}}
\thicklines
\put(0,-6){\line(1,1){20}}
\put(7,-3){\tiny $e$}
\put(-8,8){\tiny $f$}
\put(2,9){\tiny $h_{e\!f}$}
\end{picture}
\qquad \qquad      \Longrightarrow
\qquad  \qquad  
\int_{{\rm SU}(2)}dU_{e*}\; \chi^{j_f}(U_{f})\,;
\label{due}
\end{equation}

\item
We sum over each face $f*$ and associate the element
\begin{equation}
\hspace{3em}
\begin{picture}(25,25)
\thicklines
\put(-10.1,-6){\line(0,1){20}}
\put(-0.1,-6){\line(-1,0){10.3}}
\put(0,24){\line(1,0){10}}
\put(-10,14){\line(1,1){10}}
\put(20,14){\line(-1,1){10}}
\put(0,-5.9){\line(1,1){20}}
\put(11,0){\tiny $U_{e*}$}
\put(-1,8){\tiny $f*$}
\put(0,-11){\tiny $g'$}
\put(22,11){\tiny $g$}
\end{picture}
\qquad   \qquad    \Longrightarrow
\qquad  \qquad  
\sum_{j_{\!f*}}\Delta_{j_{\!f*}}\,\chi^{\scriptscriptstyle j_{\!f*}}\!\Big(\!\prod_{e*\in\partial f}U_{e*}\!\Big)\,;
\label{tre}
\end{equation}

\item
We drop, at each vertex, an integral $\int_{\rm SU(2)} dU_{v(e)}$, which appears as redundant in (\ref{uno}).

\end{itemize}

The functor $\mathcal{Z}(U_l)$ provides the transition operator between boundary states, and gives the algebraic counterpart of cobordisms between boundary manifolds. It clearly depends on the boundary group elements and it is written as
\begin{eqnarray} \label{fun}
\mathcal{Z}_\mathcal{C}(U_l)=\int_{{\rm SU}(2)^{2(E-L)-V} } dU_{v(e)} \, \int_{{\rm SU}(2)^{\mathcal{V}-L}} dU_f\, \prod_f\, \mathcal{K}_{f*}(U_{e*},U_f) \,,
\end{eqnarray}
where $\mathcal{K}_{f*}(U_{e*},U_f)$ denotes the ``face amplitude'' 
\begin{eqnarray} \label{funface}
\mathcal{K}_{f*}(U_{e*},U_f)\equiv 
\sum_{j_{f*}} \,  \Delta_{j_{f*}} \,  \chi^{\scriptscriptstyle j_{\!f*}}\!\Big(\!\prod_{e*\in\partial f}U_{s(e)}U_{e*}U^{-1}_{t(e)}\!\Big) \,  \prod_{e*\in\partial f}\!\chi^{\scriptscriptstyle j_{\!f*}}(U_f)\,.
\end{eqnarray}

Taking into account a 2-complex without boundary, (\ref{fun}) reduces to the partition function  
\begin{eqnarray} \label{fun2}
\mathcal{Z}_\mathcal{C}
&=&\int_{{\rm SU}(2)^{2E-V} } dU_{v(e)} \, \int_{{\rm SU}(2)^{\mathcal{V}}} dU_f\, \sum_{j_{f*}} \prod_f \Delta_{j_{f*}} \times \nonumber\\
&& \chi^{\scriptscriptstyle j_{\!f*}}\!\Big(\!\prod_{e*\in\partial f}(U_{s(e)}U_{e*}U^{-1}_{t(e)})\!\Big)\! \prod_{e*\in\partial f}\!\chi^{\scriptscriptstyle j_{\!f*}}(U_f)\,,
\end{eqnarray}
where $\mathcal{V}$ is the sum of the valences of the faces of $\mathcal{C}$. Differently than in (\ref{fun}), the expression in (\ref{fun2}) provides the amplitudes of probability for the output of the transition among states. This coincides to the process of ``capping'' the boundaries described before, and gives a partition function which is a topological invariant of manifolds. As observed before, for the example of TQFT in dimension $2$, this is an endomorphism of the ground field $\mathbb C$. 

We notice that the functor $\mathcal Z_{\mathcal C}$ derives its form from an integration on the possible geometries that determine a transition between boundary states. More specifically, it is known (see for example \cite{Rovelli2010} and references therein) that the partition function defined above approximates the semi-classical limit of the Einstein-Hilbert action, and the integration variables can be interpreted as living in the moduli space of (equivalent up to diffeomorphism) metrics over the base manifold. Rovelli \cite{Rovelli2010} compares this approximation to a ``concrete implementation'' of the Misner-Hawking integral. In the setting of the present article, this is interpreted as the learning rule itself. A TQNN computes transition amplitudes between states by obtaining a partition function determined by the topology of the system, and infers this by integrating over the geometries of the system, therefore selecting a geometry that optimizes the output.

We are finally able to specify the training algorithm of the model as follows.\\ 

\begin{enumerate}
\item {\bf Initialize}:\\
Associate, between boundary states that are supported on disjoint graphs\\ 
$\{ \Gamma_{\rm in},\, \Gamma_{\rm out} \, ; \partial \mathcal{C} = \Gamma_{\rm out} \cup \Gamma_{\rm in}\}$, the functorial evolution $$\mathcal{Z}_{\mathcal{C}}(\{U_l\,; l\in \mathcal{C} \}, \{\bar{j}_l\}),$$  where $ \{\bar{j}_l\}$ denote a set of parameters to be fitted in the learning process. \\   
    
\item {\bf Feedforward}:\\

{\bf 2a} compose a functor $\mathcal{Z}_{\mathcal{C}}(\{U_l\,; l\in \mathcal{C} \}, \{\bar{j}_l\})$, which is supported on a 2-complex $\mathcal{C}$, with a series of 2-complexes interpolating among either the intermediate hidden layers graphs or the boundary states' graphs. For $P$ hidden layers, labelled by $p\in P$, we have the decomposition $\mathcal{C}=\mathcal{C}_1 \dots \cup \mathcal{C}_p \cup \mathcal{C}_{P+1}$. Therefore

\begin{eqnarray}\label{eq:composition}
&&\mathcal{Z}_{\mathcal{C}}(\{U_l\,; l\in \mathcal{C} \}, \{\bar{j}_{l}\})=\\
&&\mathcal{Z}_{\mathcal{C}_1}(\{U_{l_{\rm in}}\,; {l_{\rm in}}\in \Gamma_{\rm in} \}, \{\bar{j}_{l_{\rm in}}\}) \cdots    \mathcal{Z}_{\mathcal{C}_1}(\{U_{l_{\rm out}}\,; {l_{\rm in}}\in \Gamma_{\rm out} \}, \{\bar{j}_{l_{\rm out}}\})\,, \nonumber  
\end{eqnarray}

where the dot denotes the integration over the group elements assigned to the interpolating graphs supporting the hidden layer structures. 
This, in fact, encodes functoriality of $\mathcal Z$, since it respects composition of intermediate manifolds. 

{\bf 2b} integrate over the group elements $U$ assigned to the hidden layer graphs, so to trace them out:

\begin{eqnarray} \label{cob}
\mathcal{Z}_{\mathcal{C}_1}(\{G \})\cdot  \mathcal{Z}_{\mathcal{C}_2}(\{H \})&=& 
\\
\int_{\rm SU(2)} \prod  dU\, \mathcal{Z}_{\mathcal{C}_1}(\{U \},  \{G \})\,\,  \mathcal{Z}_{\mathcal{C}_2}(\{U \}, \{H \})&=&\mathcal{Z}_{\mathcal{C}_1 \cup \mathcal{C}_2 }(\{G \}, \{H \})\,. \nonumber 
\end{eqnarray}
This property is often referred to as a cobordism of the functorial structure.\\
    
\item {\bf Classify}:\\
    
Introduce $H_l\in$ SL$(2,\mathbb{C})$, encoding the information on the set of parameters  $\{ \bar{j}_l\}$; by the aforementioned combinatorics, associate to the 2-complex $\mathcal{C}$ the transition amplitude

\begin{eqnarray} \label{funbis1a}
\mathcal{Z}_\mathcal{C}(H_l)=\int_{{\rm SU}(2)^{2(E-L)-V} } dU_{v(e)} \, \int_{{\rm SU}(2)^{\mathcal{V}-L}} dU_f\, \prod_f\, \mathcal{K}^{t_{f*}}_{f*}(U_{e*},U_f) \,,
\end{eqnarray}
where the heat kernel propagator, encoding the information about the parameter $\{\bar{j}_l \}$ through the SU$(2)$ coherent group elements \cite{Bianchi:2010mw}, acquires the expression   
\begin{eqnarray} \label{funbis2}
\mathcal{K}^{t_{f*}}_{f*}(U_{e*},U_f) &\equiv& 
\sum_{j_{f*}} \,  \Delta_{j_{f*}} \, e^{-j_{f*}(j_{f*}+1) \frac{t_{t_{f*}}}{2} } \times \nonumber\\ &&\chi^{\scriptscriptstyle j_{\!f*}}\!\Big(\!\prod_{e*\in\partial f}( U_{s(e)}U_{e*}U^{-1}_{t(e)}) H_{e*}^{-1}\!\Big) \,  \prod_{e*\in\partial f}\!\chi^{\scriptscriptstyle j_{\!f*}}(U_f)\,,
\end{eqnarray}
$\{t_{f*}\}$ being a set of positive real numbers.\\
\item {\bf Estimate}:\\
Estimate the parameters $\{ \bar{j}_l\}$, maximizing the probability derived from the amplitude $\mathcal{Z}_{\mathcal{C}}$, in a feedforward approach. \\
\item {\bf Repeat}:\\    
Repeat the previous steps 1-4 for different choices of the boundaries $\partial \mathcal{C}$.
    
\end{enumerate}

We conclude this section with few remarks about TQNN. First we notice that the definition of TQNN does not generally fix the geometry of the network, but it rather determines a ``preferred'' geometry to detect certain (equivalence classes of) states by considering the highest transition amplitudes. Moreover, implicit in the use of the transition amplitude used in loop quantum gravity again pointing at the recent discussion on mentioning LQG, we naturally implement the superposition principle, as a sum (of sorts) over all possible histories between boundary states, i.e. paths through the intervening $n$-manifold $M$. This might be compared to utilizing classical networks of arbitrary layer widths and depths simultaneously, as different histories present in general a different number of single vertex transitions that are composed to transition from one boundary state to another. Following this line of interpretation, it is reasonable to expect that ideally a TQNN ``implements all input/output equivalent DNNs in parallel'' (cf \cite{Deutsch}) and hence presents considerably higher computational performance with respect to a classical neural network. 

Interestingly, while as noted above the most straightforward interpretation of QNN as spin-networks assumes that the quantum machine corresponds to a given spin-network, in TQNN an appropriate functor determines the transition between two spin-networks that are associated to single states. This functor represents, in effect, a superposition of quantum machines implementing the chosen function $f: N_1 \mapsto N_2$ from the input to the output state.  Replacing single maps with functors representing appropriate equivalence classes of maps in this way is commonly referred to as {\it categorification} in mathematics.

\section{Associating spin-networks to images}\label{sec:Association}

A fundamental feature of the definition of TQNN is that input and output states are spin-networks and, more generally, cylindrical functions of the Hilbert space in the holomorphic representation of quantum gravity. It is therefore crucial to have well determined rules to associate spin-networks to the input data. We suppose to have a pixeled image whose shades of gray vary in $[0,1]$. We let the nodes of our spin-network coincide with the centers of the pixels. For each node $N$, we let $j_a$ denote the spin $j$ representation of $SU(2)$ where $a$ is ten times the shade of grey of the pixel whose center is $N$. Then, we consider the von Neumann neighbourhood of a node $N$, and for a node $N'$ in the neighbourhood we join the two nodes by $j_{ab} = {\rm min}\{j_a,j_b\}$, where $a$ and $b$ are the associated (re-scaled) shade of grey of the pixels of $N$ and $N'$, respectively. We apply the Jones-Wenzl projector \cite{Kauffman-Lins} to the representation corresponding to $j_{ab}$ in order to symmetrize it, so to provide all the possible spin irreducible representations with $0 \leq j \leq 5$. 

\begin{figure}
\centering
\includegraphics[width=11 cm]{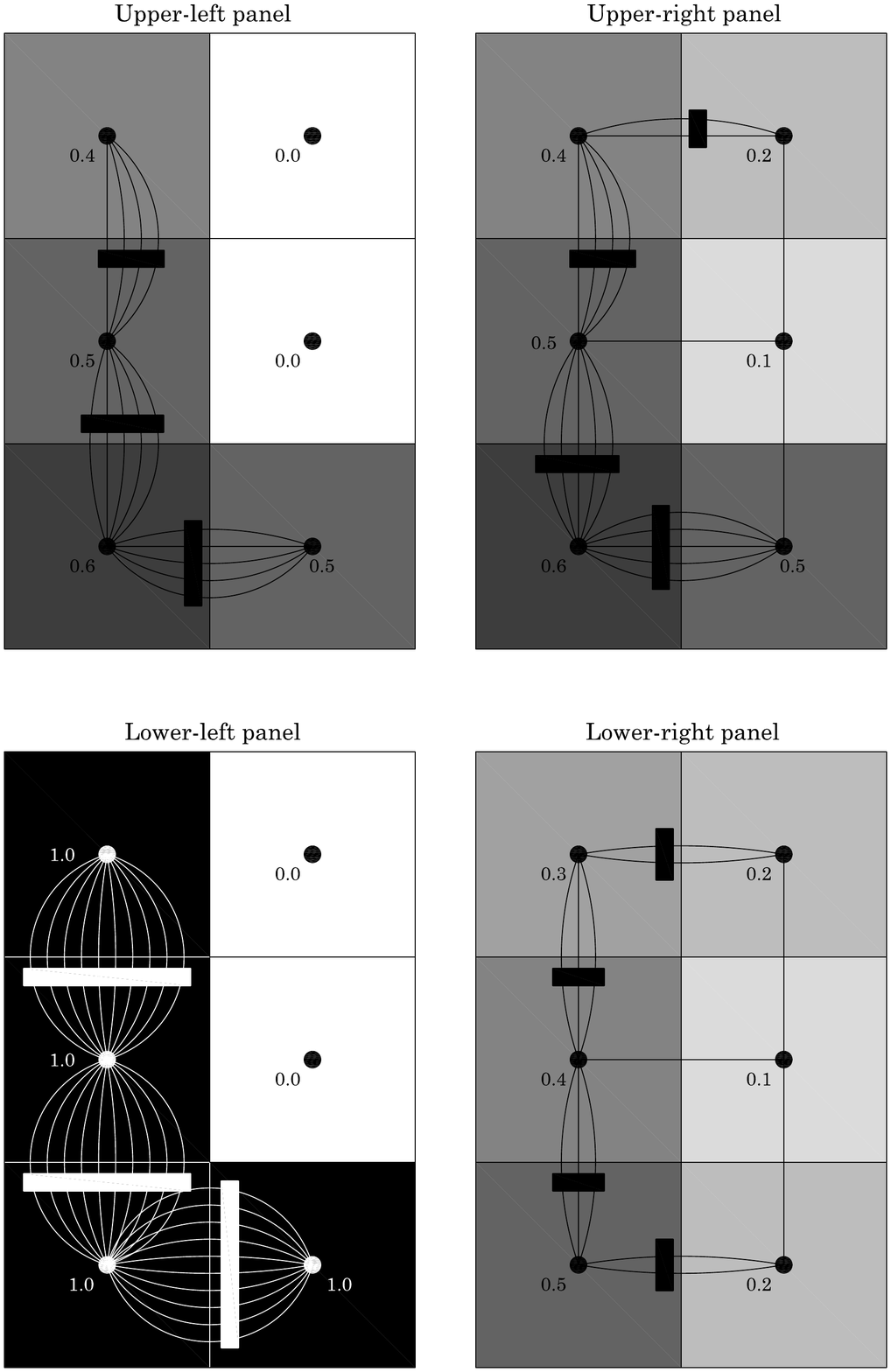}
\caption{Superimposed to four different images are the associated graphs, endowed with assigned SU$(2)$ irreducible representations. The bottom left panel encloses an image that corresponds exactly, i.e. with probability 1, to  a "L".}
\label{X}
\end{figure}

\begin{figure}
\centering
\includegraphics[width=8 cm]{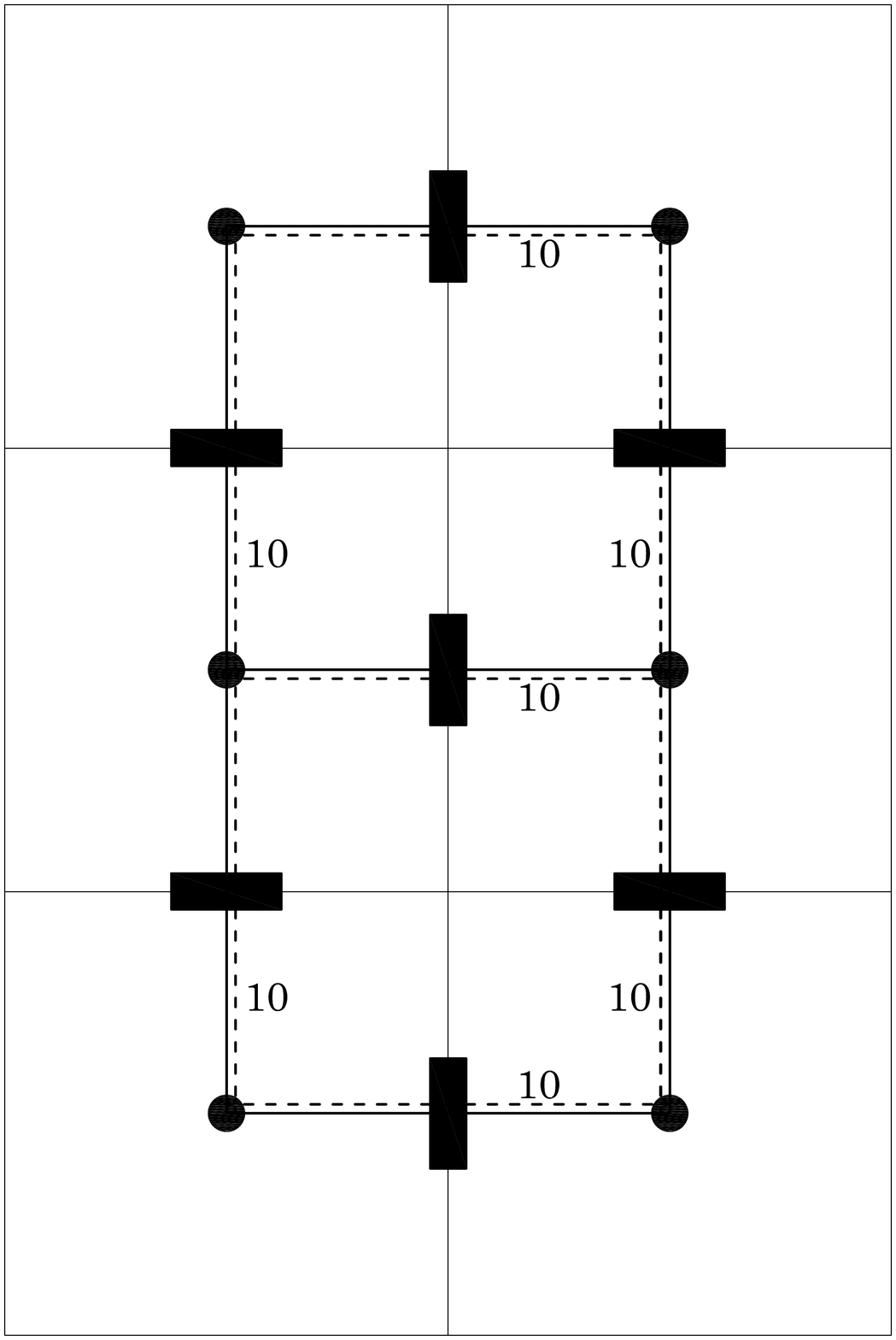}
\caption{The maximal graph, which encloses all the possible sub-graphs supporting the training samples' cylindrical functions.}
\label{XX}
\end{figure}

To better elucidate the previous scheme we consider the specific situation of handwritten letters with $3\times 2$ pixels and the shades of gray, range in the interval $[0,1]$ in decimals, where $0$ corresponds to white, while $1$ corresponds to black. By construction, the nodes of the spin-networks obtained will have $6$ nodes, each centered in one of the pixels. For example, four instances of the letter ``L'' and their corresponding spin-networks are given in Figure~\ref{X}, where we use rectangular boxes to denote the Jones-Wenzl projector applied to the edges (corresponding to $SU(2)$ representations) joining two nodes. In the case of the top left panel in Figure~\ref{X}, proceeding counterclockwise from the left top pixel, the encountered set of shades of grey is set to be $\{0.4, 0.5, 0.6, 0.5, 0, 0 \}$. A slightly different case is represented in the right up panel of Figure~\ref{X} for which the string of numbers is $\{0.4, 0.5, 0.6, 0.5, 0.1, 0.2 \}$. The ideal case, corresponding to the spin-network state that perfectly captures the letter $L$, with a probability $|\mathcal{A}|^2=1$, is given by $\{1.0, 1.0, 1.0, 1.0, 0.0, 0.0 \} \equiv L$, and is represented on the left bottom panel of Figure~\ref{X}. Finally, the left bottom panel represents an undetermined case captured by the string of numbers $\{0.3, 0.4, 0.3, 0.2, 0.1, 0.2 \}$. We denote these cases respectively as $A$, $B$, $C$ and $D$. We shall notice that these are all nothing but ``colored'' sub-graphs that can be recovered from a maximally connected graph, the one pictured in Figure~\ref{XX}, by removing fundamental representation strands along the links.   

\section{The semi-classical limit}\label{sec:Semi-Classical}

We have so far considered spin-network basis states represented by cylindrical functionals of the holonomies, contracted with the intertwiner invariant tensors. A different representation involves coherent spin-network states \cite{Bianchi:2009ky}, which is obtained as the gauge-invariant projection of the product over links of heat kernels. Namely 
\begin{eqnarray}  \label{hks}
\Psi_{\Gamma, H_{ab}}(h_{ab})=\int \left(\prod_a dg_a \right) \prod_{ab} \mathcal{K}^{t_{ab}}(h_{ab}, g_a H_{ab} g_b^{-1})\,,
\end{eqnarray}
where $a,b$ label the nodes of the maximal graph where the spin-networks live, pairs $ab$ correspond to links, $g_a\in SU(2)$ are group elements at the nodes, $h_{ab}\in SU(2)$ label group elements over the links, and $H_{ab}$ are group elements of SL$(2,\mathbb{C})$, assigned to each link $ab$. Notice that elements of  SL$(2,\mathbb{C})$ can be expressed in terms of a positive real number $\eta_{ab}$ and two independent SU$(2)$ group-element $g_{ab}$ and $g_{ab}^{-1}$, namely
\begin{equation}
    H_{ab}=g_{ab}e^{\eta_{ab}(\sigma_3/2)}g_{ba}^{-1}\,.
\end{equation}
The two SU$(2)$ group elements cast uniquely in terms of an angle $\tilde{\phi}$ and a unit vector identified by its inclination and azimuth $\vec{n}=(\sin \theta \cos \phi, \sin \theta \sin \phi, \cos \theta )$. The associated SU$(2)$ group element reads 
\begin{equation} \label{rota}
n=\exp (-\imath \phi \sigma_3/2) \exp (-\imath \theta \sigma_2/2))\,,
\end{equation}
and the SU$(2)$ group elements $g$ recast $g=n \exp (\imath \tilde{\phi}\sigma_3/2)$. Thus we get
\begin{equation}
    H_{ab}=n_{ab}e^{-\imath z_{ab}(\sigma_3/2)}n_{ba}^{-1}\,.
\end{equation}
having introduced $z_{ab}=\xi_{ab}+\imath \eta_{ab}$, with $\xi_{ab}=\tilde{\phi}_{ba}-\tilde{\phi}_{ab}$. This finally allows to identify the set of parameters associated to each link, namely $(\vec{n}_{ab}, \vec{n}_{ba},\xi_{ab},\eta_{ab})$. 
These parameters give weight vectors that determine the transition amplitudes that the TQNN associates to input and output states. The learning process, therefore, consists of obtaining the weights that produce the maximal transition amplitudes with respect to a ground truth. For example, in the case of spin-networks associated to handwritten letters ``L'' given above, the weights have to maximise the transition amplitude corresponding to the lower bottom panel of Figure~\ref{X}. 

The state in Eq.~\eqref{hks} can be expanded on the spin-network basis $\Psi_{\Gamma, j_{ab}, \iota_a}$, 
\begin{eqnarray}
\Psi_{\Gamma, H_{ab}}(h_{ab})=\sum_{j_{ab}} \sum_{\iota_a} \, f_{j_{ab}, \iota_a}\, \Psi_{\Gamma, j_{ab}, \iota_a}(h_{ab})
\,,
\end{eqnarray}
with coefficients $f_{j_{ab}, \iota_a}$ individuated by  
\begin{eqnarray}
f_{j_{ab}, \iota_a}= \left( \prod_{ab} \Delta_{j_{ab}}\, e^{-j_{ab} (j_{ab}+1 )t_{ab}} D^{j_{ab}}(H_{ab}) \right) \cdot \left(\prod_a v_{\iota_a} \right) \,.
\end{eqnarray}
In the large $j_{ab}$ limit, the coherent states $\Psi_{\Gamma, H_{ab}}(h_{ab})$ undergo the expansion 
\begin{eqnarray}\label{eq:semiclassicalpsi}
\Psi_{\Gamma, H_{ab}}(h_{ab})\simeq \sum_{j_{ab}} \left( \prod_{ab} \Delta_{j_{ab}}\,  e^{-\frac{\left( j_{ab}- \bar{j}_{ab} \right)^2}{2 \sigma_{ab}^2 }  } \, e^{-\imath \xi_{ab} j_{ab}} \right)\, \Psi_{\Gamma, j_{ab}, \Phi_{a}(\vec{n}_{ab})} (h_{ab}) \,,
\end{eqnarray}
where the coherent intertwiners $\Phi_a(\vec{n}_{ab})$ can be decomposed on the intertwiner space $v_{\iota_a}$ by
\begin{eqnarray}
\Phi_a(\vec{n}_{ab})= \sum_{\iota_a} \Phi_{\iota_a}(\vec{n}_{ab}) v_{\iota_a}\,,
\end{eqnarray}
with 
\begin{equation}
\Phi_{\iota_{a}}(\vec{n}_{ab})= v_{\iota_{a}} \cdot \left( \bigotimes \limits_b | j_{ab},\vec{n}_{ab} \rangle \right)\,,
\end{equation}
the variance of the Gaussian distribution per each link is inversely proportional to the diffusion time $t_{ab}$, namely  $\sigma_{ab}\equiv 1/(2\, t_{ab}) $, and finally the parameters $\bar{j}_{ab}$ over which the coherent state is peaked, which correspond to the estimated parameters we refer to through the paper, are related to the $\eta_{ab}$, the real numbers entering the parametrization of SL$(2,\mathbb{C})$ group elements, at each link by $\Delta_{\bar{j}_{ab}}\equiv \eta_{ab}/t_{ab}$. \\

The partition function of Section~\ref{sec:TQNN} is therefore changed in the semi-classical limit by the use of the approximations in Eq.(\ref{eq:semiclassicalpsi}) and the corresponding transition amplitudes between a initial and final states $\Psi_{\Gamma, {j}_\gamma, \iota_n}, \Psi_{\Gamma, H_{ab}}$, respectively, are therefore computed according to the formula:

\begin{eqnarray} \label{Ax}
\mathcal{A}_{\prod_{ab} H_{ab}}&=& \langle 
\Psi_{\Gamma, H_{ab}} |  \Psi_{\Gamma, {j}_\gamma, \iota_n }  \rangle \simeq  \sum_{j_{ab}} \left( \prod_{ab} \Delta_{j_{ab}}\,  e^{-\frac{\left( j_{ab}- \bar{j}_{ab} \right)^2}{2 \sigma_{ab}^2 }  } \, e^{-\imath \xi_{ab} j_{ab}} \right)\, \nonumber\\
&\phantom{a}& \times \int dh_{ab}  
\overline{\Psi}_{\Gamma, j_{ab}, \Phi_{a}(\vec{n}_{ab})} (h_{ab}) \Psi_{\Gamma', {j'}_{ab}, v_{{\iota'}_a}} (h_{ab}) \nonumber\\
&=& \sum_{j_{ab}} \left( \prod_{ab} \Delta_{j_{ab}}\,  e^{-\frac{\left( j_{ab}- \bar{j}_{ab} \right)^2}{2 \sigma_{ab}^2 }  } \, e^{-\imath \xi_{ab} j_{ab}} \right) \delta_{\Phi_{a}(\vec{n}_{ab}), v_{{\iota'}_a}} \delta_{j_{ab} {j'}_{ab}} \nonumber\\
&=& \left( \prod_{ab} \Delta_{j_{ab}}\,  e^{-\frac{\left( j_{ab}- \bar{j}_{ab} \right)^2}{2 \sigma_{ab}^2 }  } \, e^{-\imath \xi_{ab} j_{ab}} \right) 
\,.
\end{eqnarray}
Using the transition amplitudes above, between states in the semi-classical limit, we can apply the fundamental idea of the  algorithm of Section~\ref{sec:TQNN} in the semi-classical limit to obtain:

\begin{enumerate}
    
    \item {\bf Initialize}:\\
    
   Associate spin-networks to images as in Section~\ref{sec:Association}. This is done in two steps:  \\
  
{\bf 1a} associate to each training sample a 1-complex (i.e. a graph), where each node corresponds to the center of a pixel, and the edges connect pixels in the von Neumann neighbourhoods; \\

{\bf 1b} assign to each link of the 1-complex SU$(2)$ irreducible representations, where the spin $j$ representation label is determined by the pixel colours. \\

    \item {\bf Feedforward}:\\

{\bf 2a} estimate the parameters entering the feedforward pattern through the functorial functional $\mathcal{Z}_{\mathcal{C}}(h_l)$, by maximizing the internal product $\mathcal{A}$ between this latter and the QNN boundary states supported on $\partial \mathcal{C}$. The geoemtric supports for QNN boundary states are graphs resulting from the disjoint union of any $\Gamma'$, on which training samples are constructed, and 1-complexes supporting output states; \\

{\bf 2b} for hidden layer approaches: compute the functorial composition (cobordism properties) to take place accordingly to Eq.~(\ref{cob}), and consistently with the filtering process that is implemented by the selection of the sub-graph structure at each hidden layer.\\
    
\item {\bf Classify}:\\
    
{\bf 3a} introduce $H_l\in$ SL$(2,\mathbb{C})$, encoding the information on the set of parameters to be determined, namely $(\vec{n}_{ab}, \vec{n}_{ba},\xi_{ab},\eta_{ab})$; \\

{\bf 3b} associate to each link of the  1-complex a set of parameters, the string $(\vec{n}_{ab}, \vec{n}_{ba},\xi_{ab},\eta_{ab})$, to be fitted in the learning process. This identifies the functional $\Psi_{\Gamma, H_{ab}}$;\\

{\bf 3c} compute the internal product to associate probability amplitudes to the training samples:
\begin{equation}
\mathcal{A}_{\prod_{ab} H_{ab}} = \langle 
\Psi_{\Gamma, H_{ab}} |  \widetilde{\Psi}_{\Gamma, {j}_\gamma, \iota_n }  \rangle\,,
\end{equation}
the $\Psi_{\Gamma, H_{ab}}$ denoting the functionals of the training samples, and $ \widetilde{\Psi}_{\Gamma, {j}_\gamma, \iota_n }$ the functional associated to the image to be recognized.\\

\item {\bf Estimate}:\\

 Estimate, for each training sample, the parameters $(\vec{n}_{ab}, \vec{n}_{ba},\xi_{ab},\eta_{ab})$, maximizing the probability derived from the amplitude $\mathcal{A}_{\prod_{ab} H_{ab}}$.\\

These parameters individuate a rotation group element Eq.~(\ref{rota}), which acting on a reference vector, e.g. the identity element of the SU$(2)$ group, individuates the weight vector.
\\
    
\item {\bf Repeat}:\\ 

Repeat the previous steps for different cylindrical functions, corresponding to different training samples, by using the estimated parameters, and the corresponding weight vectors.
\end{enumerate}

Observe that the topological structure of the graph, and the related extended information that is encoded by its links and intertwiners, are captured by the combinatorial summation of the $a,b$ indices, and by the information stored in the Kronecker delta on the projected coherent intertwiners at each node. On the other hand, metric properties are encoded in the Gaussian weights at each link, capturing the relevant quantitative information concerning the recognition of the specific digit. It is clear that the case in which, at the link $\gamma_{ab}$, both the mean value $\bar{j}_{ab}$ and its dispersion $({j}_{ab}-\bar{j}_{ab})^2/\sigma^2_{ab}$ are vanishing, no information relative to that link appears anymore in the amplitude, and the specific metric feature affects the topology of the graph, with the consequence that the graph will embed one link less. Finally, we recognize as a remarkable feature of this approach that probability interference terms (while computing $|\mathcal{A}|^2$) will be provided by the $\xi_{ab}$ coefficients.

\subsection{The perceptron in the semi-classical limit}

We consider now our topological version of the notion of perceptron, and show that in the semi-classical limit we obtain an object that resembles traditional perceptrons closely. The first step toward adapting TQNN to the setting of perceptrons, is to define an algorithmic way to associate spin-networks to input vectors in $\mathbb R^n$, that constitute the dataset. Let $N$ be a natural number which is large compared to the magnitudes of the entries of the vectors of the dataset. Given a vector $\bar x$, we construct a spin network $\Gamma_{\bar x}$ associated to $\bar x$ as follows. We introduce a node which is labeled by $0$, and for each $i = 1, 2, \ldots , n$ we add a node, labeled by the index $i$ of the corresponding entry of $\bar x$. As in the case of Section~\ref{sec:Association}, we colour the node labeled by $0$ with the spin representation $j_N$, while each node $i$ is coloured by $[x_i]$, the closest integer rounding $x_i$. Then, for each $i$ we inroduce an edge connecting $0$ and $i$, which is labeled by a spin $j_{0i} = N + [x_i]$ representation. Finally, we symmetrize the edges by applying the Jones-Wenzl projector, indicated diagrammatically by placing a black box on the connecting edges. Observe that we do not introduce, in this context, links between nodes $i$ and $j$ with $i,j \neq 0$. Now, the weights of the perceptron are vectors $\bar w\in \mathbb R^n$ similarly to the inputs $\bar x$ of the dataset. We follow the same procedure above to introduce a spin-network $\Gamma_{\bar w}$ of weights.

Since we have chosen $N$ much larger than the actual range of the data entries $\bar x$ (i.e. the hypercube $[-M,M]^n$ where $M$ is the maximum magnitude that the entries of the dataset reach, has $M<< N$), it follows that we can adopt the large spin $j_{0i}$ limit, for which transition amplitudes are computed as
\begin{eqnarray}
\mathcal A_{\prod_{i} H_{0i},\bar w} &=& \langle \Psi_{\Gamma_{\bar x},H_{0i}}  | \Psi_{\Gamma_{\bar w}, j_{\bar w}, \iota_n}\rangle 
= \prod_{i} \Delta_{j_{0i}}\,  e^{-\frac{\left( j_{0i}- \bar{j}_{0i} \right)^2}{2 \sigma_{0i}^2 }  } \, e^{-\imath \xi_{0i} j_{0i}}.
\end{eqnarray}
The analogy with classical perceptrons is as follows. A perceptron trains a function $f$ whose weight vector $\bar w$ determines the output according to the rule $f(\bar x) = 1,0$ depending on whether $\bar w\cdot \bar x > \theta$ or not, respectively, for some threshold $\theta$, and where $\cdot$ indicates the inner product of $\mathbb R^n$. In fact, usually a bias appears in the perceptron formulas, but this can be encoded among the weights as well, so we will omit referring to it. In our topological version above, the amplitude $\mathcal A_{\prod_{i} H_{0i},\bar w}$ is obtained by the inner product of spin-network states associated to inputs $\bar x$ and weights $\bar w$. The transition amplitude $\mathcal A_{\prod_{i} H_{0i},\bar w}$ is a complex number whose modulus square is between $0$ and $1$, so that by applying a Heaviside step function $H$, centered at some threshold value $\theta$, to $|\mathcal A_{\prod_{i} H_{0i},\bar w}|^2$ we obtain a TQNN implementation of the concept of perceptron. Training a topological perceptron would account to optimizing weights $\bar w$, and $SL(2,\mathbb C$) elements $H_{0i}$ with respect to a predetermined target. 

A similar reasoning applied to feedforward neural networks (i.e. multilayer perceptrons) can be implemented as well, by using the fact that TQFTs are defined via functorial constructions that allow us to compose an arbitrary number of computational units as above.  Note that in this setting the ``semi-classical'' nature of QNNs with fixed layers and fixed connections, and hence classical constraints on entanglement between qubits, also becomes clear: such systems effectively choose only particular paths through the input/output equivalent TQNN to implement, enforcing this choice architecturally. We see, therefore, that TQNNs are versatile objects that can be trained and utilized for classification problems in different ways. Moreover, through the notion of semi-classical limit, they provide a way of interpreting artificial neural networks in the context of TQNN theory. 

\section{Experiments on handwritten letter recognition}\label{sec:Example}

We consider now the theory introduced in this article, applied to a concrete example. It is worth mentioning that we take into account hidden layers, i.e. 2b in the ``Feedforward'' step of the algorithm of Section~\ref{sec:Semi-Classical}. This consists of interpolating among intermediate states, on which a complete summation is taken into account through Eq.~(\ref{cob}), and which are supported only on a restricted set of sub-graphs. The functoriality of TQNN in this sense is here fundamental, as Eq.~(\ref{cob}) encodes precisely the composition property of cobordisms, preserved by topological quantum field theories. We can imagine the hidden layers to act as filtering specific patterns over others. Indeed, what the hidden layers do is to impose a selection over the intermediate graphs $\partial \mathcal{C}_n$, and hence the 2-complexes that interpolate among these latter ones. Internal summation over the irreducible representations of SU$(2)$, namely variation of the metric properties of the QNN states, then individuates all the possible sub-graphs contained in $\partial \mathcal{C}_n$, i.e. corresponds to a variation of the topological features of the 1- and 2-complex structures.

Applying the definition of cobordisms and functoriality implicit in the definition of TQNN as a type of TQFT, implementing different layers as described above simply coincides with computing transition amplitudes through middle steps in the computation, as prescribed by Eq.~(\ref{eq:composition}). 

\begin{figure}
\centering
\includegraphics[width=12 cm]{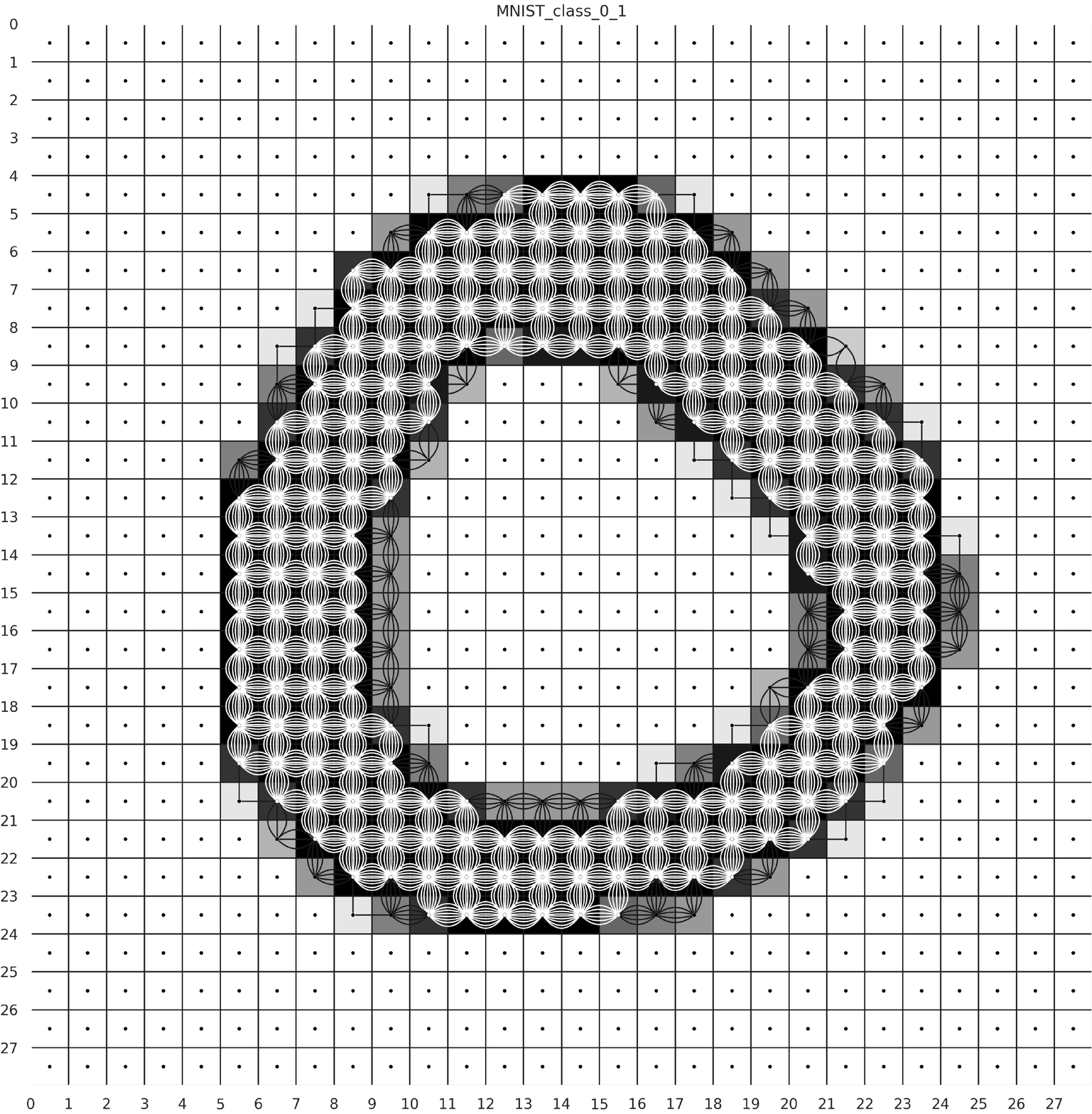}
\caption{A specific graph, representing a the number $0$, within the case employing $28 \times 28$ pixels.}
\end{figure}

\begin{figure}
\centering
\includegraphics[width=12 cm]{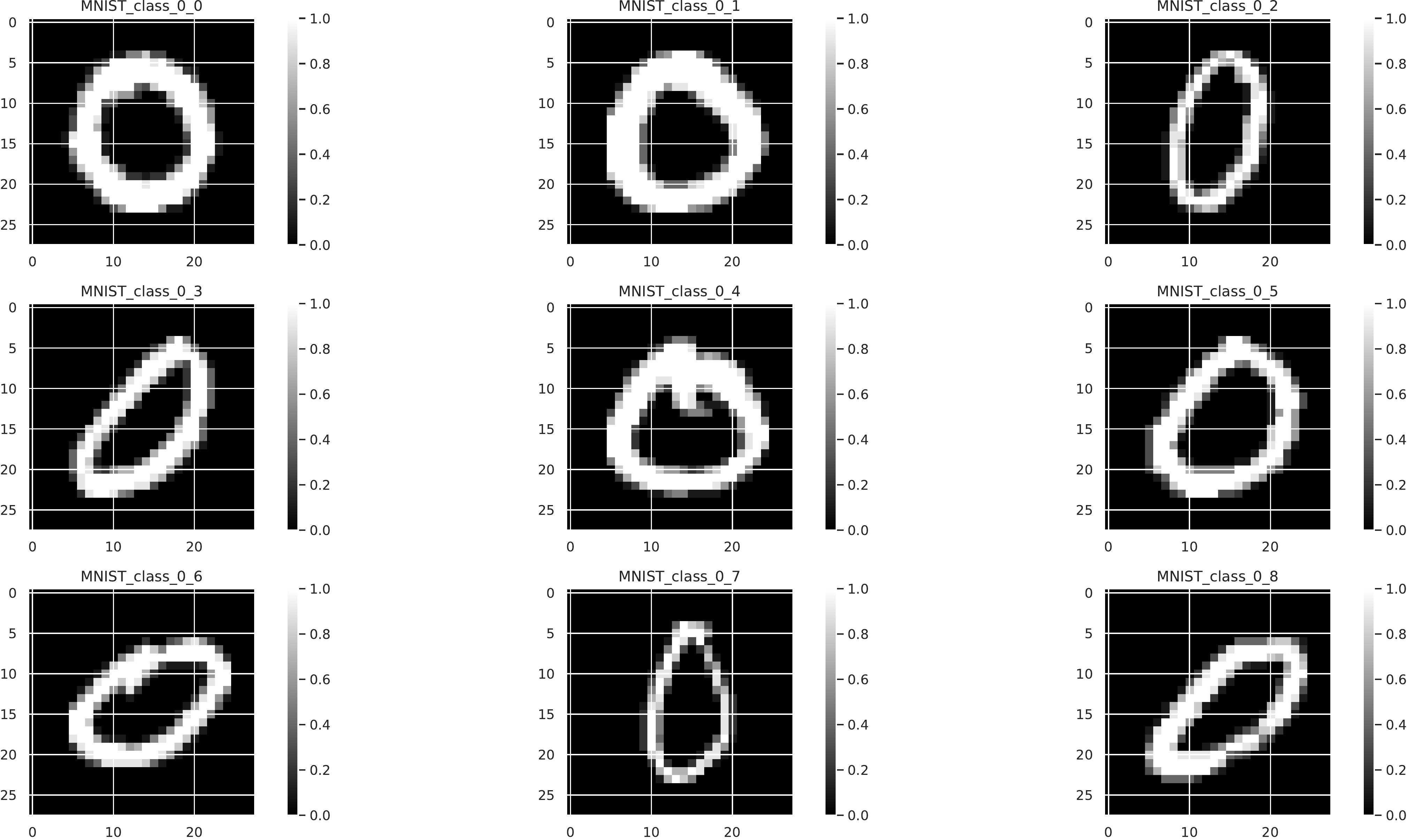}
\caption{Several samples of the number $0$, extracted from the MNIST data base, to be used during the training process.}
\label{MNISTdata}
\end{figure}

\begin{figure}
\centering
\includegraphics[width=5 cm]{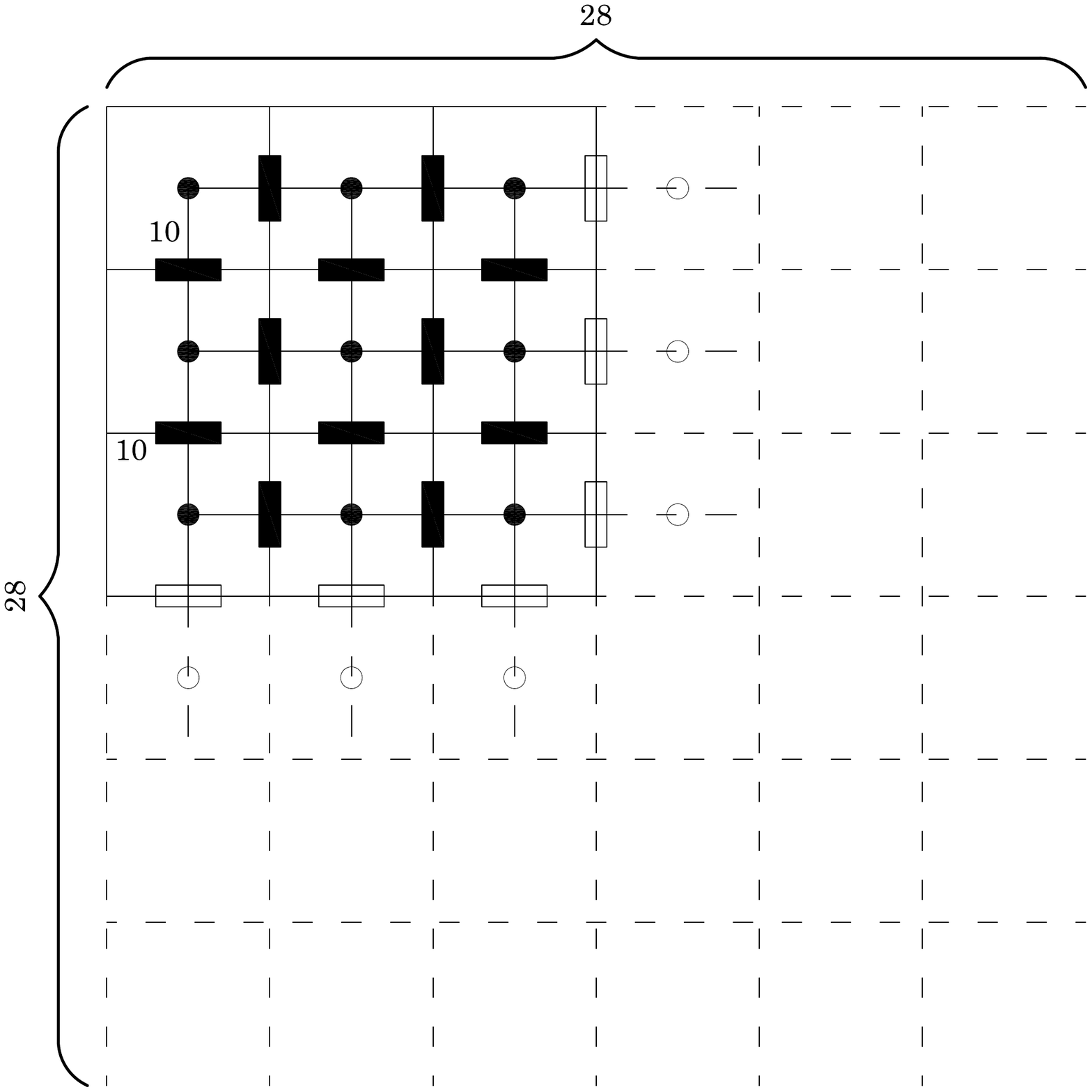}
\caption{The maximal graph, which encloses all the possible sub-graphs supporting the training samples’ cylindrical functions for the case $28 \times 28$ pixels.}
\label{pixels}
\end{figure}

\begin{figure}
\centering
\includegraphics[width=12 cm]{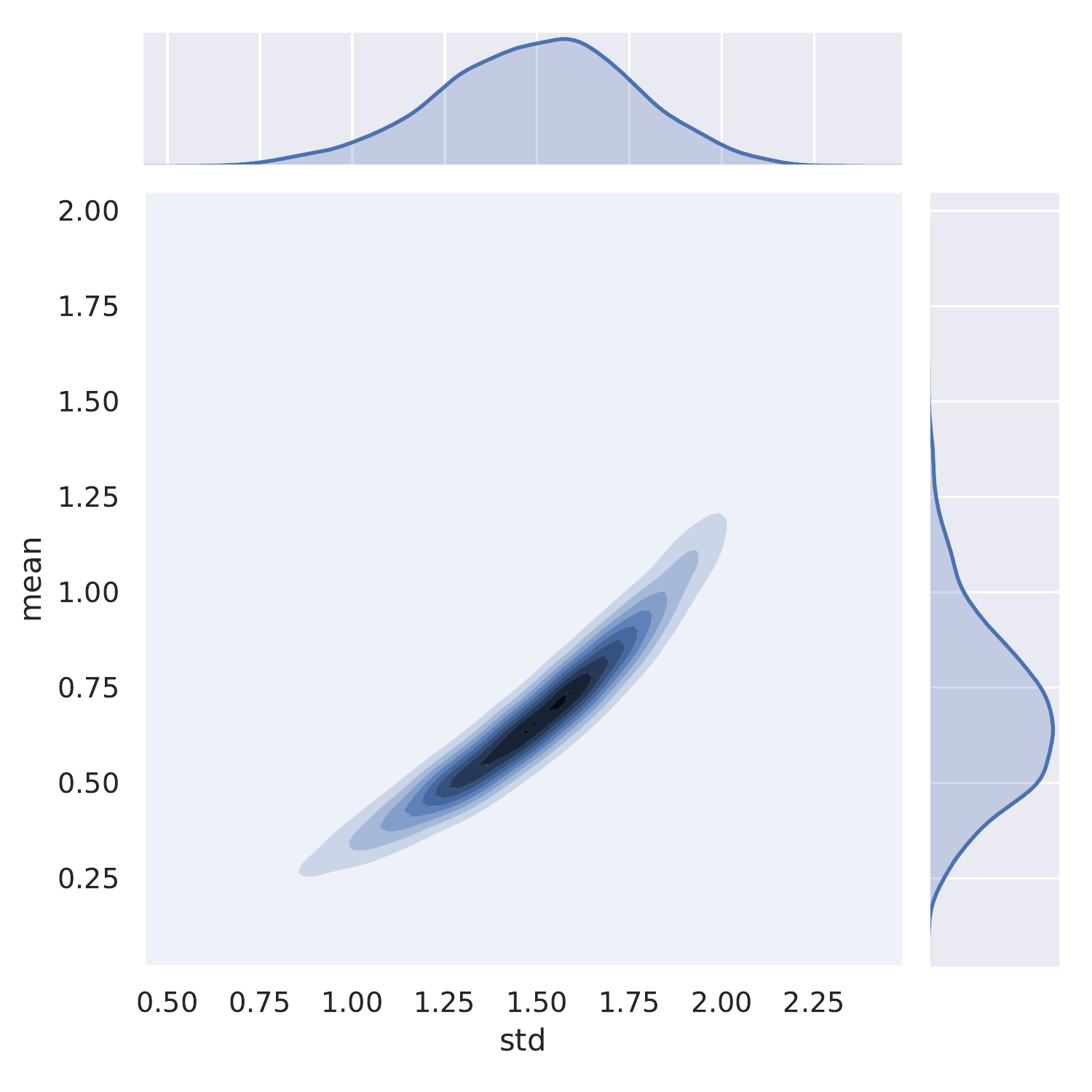}
\caption{Marginalised plots for the estimated mean values and standard deviation of the irreducible representations associated to the links of the spin-networks states.}
\label{mean}
\end{figure}

 The experiment utilizes the MNIST database (Figure~\ref{MNISTdata}) which is the standard computer vision benchmark for hand-written digit recognition. The data set contain the grey-scale image of hand-written digit. The fact that all images in the dataset have identical dimensions, which is 28 x 28 pixels, see Figure~\ref{pixels}, implies that the knowledge representation graph can be constructed from any image in the dataset. After the translation of knowledge representation graph, the parameters for each digit class are obtained using class prototyping. This consists of averaging the spin colours appearing in the training set of MNIST, in order to determine a representative spin-network whose transition with respect to input data provides the classification probability (hence the label).
The topological forms of spin-network are encoded in parameters which determine the likelihood of spin-networks state as a class. Alternatively, any optimization technique like gradient descent can be applied to learning the class prototype of specific set of spin-networks state.

The transition amplitudes are computed in the semi-classical limit using the formulas described in Section~\ref{sec:Semi-Classical}, through the implementation of the pseudo-algorithm thereby provided. In Figure~\ref{mean} we report the mean values of the standard deviations of the $j$-spin colourings corresponding to irreducible representations associated to the spin-networks. 

An implementation of TQNN without employing the semi-classical limit will appear elsewhere. Such an algorithm utilizes the machinery of Section~\ref{sec:TQNN} in its generality. We limit ourselves to mentioning that transition amplitudes, in the general setting, use the definition of Jones-Wenzl projector at the links of spin-networks, along with the projector of Noui and Perez (\cite{Noui-Perez}) to regularize the inner products.

\section{A dictionary for Quantum Neural Networks} 

As we have already mentioned, the novelty of our model consists in using the richer structures of graph-supported spin-network states to represent training and test samples. As a matter of fact, as far as we know, it is the first time that graph structures are taken into account, together with their evolution supported on 2-complexes. Instead, within the traditional approach, nodes that are located at each boundary and hidden layer, are taken to evolve along graphs (1-complexes). 

Now we are ready to reformulate notions found in DNN theory in the language of TQNN. We restrict our illustration to the supervised learning scenario consisting, as it is well known, in learning a (typically unknown) function $g: X \rightarrow Y$ that maps a (typically large, e.g. all possible images of handwritten characters) input set $X$ to a (typically much smaller, e.g. names of characters) output set $Y$, based on a training set $X^{\prime} \subset X$ and hence an explicitly represented function $g^{\prime} : X^{\prime} \rightarrow Y$ specifying example input-output pairs.  If $f: X \rightarrow Y$ is the (presumably random) function implemented by the network before training, we can represent the learning algorithm as an operation $\mathcal{L}: (f, g^{\prime}) \mapsto g$ on the initial function $f$ given the training function $g^{\prime}$. In particular, we follow the statistical learning framework of supervised learning delineated in \cite{shalev-shwartz}. Let us recall first, some classical definitions for DNN, see \cite{shalev-shwartz}. 
\begin{itemize}
\item
Sample complexity:\\
It represents the number of training-samples (i.e. $Card(X^{\prime})$) that a learning algorithm needs in order to learn successfully a family of target functions.

\item
Model capacity:\\
It is the ability of the model to fit a wide variety of functions; in particular, it specifies the class of functions $\mathfrak{H}$ (the hypothesis class) from which the learning algorithm $\mathcal{L}$ can choose the specific function \emph{$\mathfrak{h}$}.

\item
Overfitting:\\
A model is overfitting when the gap between training error and test error is too large; this phenomenon occurs when the model learns the training function $g^{\prime}$ but $\mathcal{L}$ incorrectly maps $(f, g^{\prime}) \mapsto h \neq g$, i.e. the trained network generalizes to the wrong function $h$ and fails to predict future observations (i.e. additional sample from $X$) reliably.  The training function $g^{\prime}$ has been merely ``memorized'' to the extent that $h$ is random on $X$ outside of the training sample $X^{\prime}$.

\item
Underfitting:\\
A model is underfitting when it is not able to achieve a sufficiently low error on the training function $g^{\prime}$; this phenomenon occurs when the model does not adequately capture the underlying structure of the training data set and, therefore, may also fail to predict future observations reliably. 

\item
Bias:\\
It is the restriction of the learning system towards choosing a classifier or predictor \emph{$\mathfrak{h}$} from a specific class of functions $\mathfrak{H}$ (the hypothesis class).

\item
Empirical Risk Minimization (ERM):\\
It consists in minimizing the error on the set of training data (the ``empirical" risk), with the hope that the training data is enough representative of the real distribution (the ``true" risk).

\item
Generalization:\\
It is conceived as the ability of the learner to find a predictor, i.e. a map $X^{\prime} \rightarrow X$, which is able to enlarge successfully its own predictions from the training samples to the test or unseen samples.

\end{itemize}

These notions can be translated into the TQNN dictionary as follows:

\begin{itemize}

\item 
Sample complexity:\\
It is a measure of the Hilbert-space of the entire spin-network state that is supported on a specific graph $\Gamma$. It is then dependent on the connectivity of the graph (nodes and links of each graph, i.e. the multiplicity of connectivity that characterizes the graph $\Gamma$) and on the dimensionality of the Hilbert spaces connected to each link and node. In this sense complexity, once extended to the different classes of graphs corresponding to the training set, provides a measure of the entropy of the set. Therefore, in the TQNN framework, the notion of “complexity” has a wider meaning than its counterpart in DNN, for which the sample complexity is nothing but the size of the training set. This is summarized in the expression for the dimension of the Hilbert space $\mathcal{H}_{\Gamma}$ of the (whole) spin-network supported on $\Gamma$, namely 
$$
{\rm dim}[\mathcal{H}_{\Gamma}]=\oplus_{j_l} \otimes_n \otimes_{l\in \partial n} \, {\rm dim}[\mathcal{H}_{j_l}].
$$
This directly encodes both the size of the maximal graph where the input/output states live, as well as the algebro/analytical structure used in the TQFT from which the corresponding TQNN arises, as encoded by the dimensionality of the Hilbert spaces $\mathcal H_j$, for instance;

\item
Model capacity:\\
It is quantified in terms of the interconnectivity of the graph $\Gamma$. It depends on the topological structure of the graphic support $\Gamma$ of the spin-network states, and neither on the dimensionality of the Hilbert space of the irreducible representations nor on the intertwiner quantum numbers, respectively assigned to each link and node of $\Gamma$; in other words, it depends on the total valence $V$ of $\Gamma$, defined in terms of the valences $v_n$ of each node of $\Gamma$ through the expression
$$
V=\sum_n v_n \, ;
$$
\item
Overfitting:\\
As pointed out in Section~\ref{sec:TQNN}, in the semi-classical limit, the integrals that allow us to compute the transition amplitudes that characterize a TQFT are interpreted as a ``sum over all the geometries'' of the ground topological manifold, where the integrand is some approximation of the Einstein-Hilbert action. During the learning process, then a TQNN learns how to select certain geometries with respect certain others in order to maximise certain transition amplitudes corresponding to ``a more suitable'' classification. The information available to make this selection during the learning process is that given by the connectivity of the input graphs/spin-networks and their given correlation $g^{\prime}$ with the label set $Y$. If $g^{\prime}$ is insufficiently representative of the target function $g$, the TQNN may only partially capture the topological structure of the full input set $X$ and therefore be unlikely to classify correctly spin-network states that are not part of, or are significant dissimilar from those contained in, the training set $X^{\prime}$;

\item
Underfitting:\\
It represents the converse of the overfitting scenario. The geometries that have been selected in the learning process do not correspond to the graphs $\Gamma$ at the starting point. Less information channels (links) are present, and lower dimensionality of the information channels (dimensions of the Hilbert space associated to each holonomy) as well. As a consequence, the QNN cannot fit the training set and may therefore also fail to predict future observations reliably;
 
 \item
Bias:\\
It amounts to the predisposition of the spin-network to account for a specific set of data; it depends on the topological structure of the spin-network states, encoded in the connectivity properties of input $\Gamma$'s and on the specific realization of the TQNN quantum state, i.e. on the weight of the quantum state on the spin-networks basis elements of the Hilbert space.

\item
Empirical Risk Minimization (ERM):\\
It is the variance of the Gaussian distribution of the irreducible representations assigned to the holonomies on the links in the semi-classical limit, i.e.
$$
{\rm ERM}:= \sum_l \frac{(j_l -\bar{j}_l)^2}{2 L}
\,,
$$
with $L$ equal to the total number of links.

\item
Generalization:\\
It is the behavior of the system in response to test or unseen data analogous to a functor (amplitude) either from a boundary spin-network to another boundary spin-network, or from a boundary spin-network to a complex number.
This is determined by the geometries that have been selected as the most representative of a certain training sample during the learning process. This is in practice captured by the parameters that give higher relevance, in the integral computing the transition amplitudes in a TQNN, to certain boundary transitions, while suppress others. These parameters are determined by (i) connectivity of 1- and 2-complexes (nodes and links, vertices and edges respectively), (ii) linking and knotting (e.g. for loops in a different Hilbert space representation), and (iii) states’ sum (as a global topological charge, invariant under refinement of the triangulation, i.e. invariant under refinement of the data/group elements/intertwiners assigned to the links and the nodes). How the parameters determine the corresponding amplitudes is clear, for the TQNN used in practice in this article, from the formula for the partition function of the model:

\begin{eqnarray} \label{funbis1}
\mathcal{Z}_\mathcal{C}(U_l)=\int_{{\rm SU}(2)^{2(E-L)-V} } dU_{v(e)} \, \int_{{\rm SU}(2)^{\mathcal{V}-L}} dU_f\, \prod_f\, \mathcal{K}_{f*}(U_{e*},U_f) \,,
\end{eqnarray}
where the ``face amplitude'' casts 
\begin{eqnarray} \label{funbis3}
\mathcal{K}_{f*}(U_{e*},U_f)\equiv 
\sum_{j_{f*}} \,  \Delta_{j_{f*}} \,  \chi^{\scriptscriptstyle j_{\!f*}}\!\Big(\!\prod_{e*\in\partial f}U_{e*}\!\Big) \,  \prod_{e*\in\partial f}\!\chi^{\scriptscriptstyle j_{\!f*}}(U_f)\,.
\end{eqnarray}

\end{itemize}

 Finally, from the definitions of the present article, we can provide the meaning of Learner's input and output in the context of TQNN.

\begin{itemize}
\item
Learner’s input:\\
i) The domain set X: It corresponds to links $l$ and nodes $n$, and attached holonomies $U_l$ and invariant tensors $\iota_n$ respectively along the links and at the nodes: it is concisely denoted as a state of the Hilbert space of the theory:   
$$
\Psi_{\Gamma ; \{j_l\}, \{\iota_n\}}[A] \equiv \Psi_\Gamma(U_l, \iota_n) := | \Gamma; \{j_l\}, \{\iota_n\}  \rangle ;    
$$
ii) The label set Y: It is a set of topological charges and quantum numbers, with which the 2-complex is endowed; for instance, recalling the group-isomorphism $\pi_3(S_3)$, for the mapping individuated by the homotopy group  $\pi_3(S_3)=\mathbb{Z}$ the winding number $w$ is defined as the integral over the SU$(2)$ group element 
$$
w=\frac{1}{24 \pi^2} \int_{\rm SU(2)} dU ;    
$$
\\
iii) The training data S: It is the union of the (initial) boundary colored graphs together with the topological invariants associated to them through the QNN functorial action.

\item
Learner’s output:\\
It is a prediction rule, i.e. the QNN functor that identifies the topological charges of the boundary states (training/test samples) and thus implements the classifier; for $\Gamma$ supporting a disjoint boundary state, the classifier is captured by the probability amplitude that results from the internal product 
$$
 \mathcal{A}=\langle  \Gamma; \{j_l\}, \{\iota_n\} | \, | \mathcal{Z}_{\mathcal{C}, \partial \mathcal{C}=\Gamma } ;  \{j_l\}, \{\iota_n\}\rangle \,;    
$$
\end{itemize}

\section{The notion of generalization in DNN and TQNN}

Let us now consider in detail the issue of generalization in TQNN, and a consequent attempt at answering the problem raised in \cite{zhang} for DNN.

 Firstly, let us describe the notion of randomization of the labels in the training set, in the context of TQNN. Specifically, this is when labels are generated with an approximately flat spectrum on the initial spin-network states. This corresponds to the selection of one element of the Hilbert space, with random assignment of labels, which therefore represent a natural definition of randomizing the labels in the training set.\\

We argue that the problem formulated in \cite{zhang} finds a natural explanation to the extent that we enlarge DNN into the richer structure of TQNN (supported on graphs and endowed with topological ``storage'' capabilities) and understand the traditional DNN architectures as the semi-classical limit of the TQNN counterparts. In brief, a classical DNN has only the function $g^{\prime}$ to learn; it has no access to the ``intrinsic'' structure of the training examples.  TQNN, however, are sensitive to such intrinsic structure in the form of topological invariants. Since we are addressing the generalization problem in the DNN framework from the TQNN side, we shall consider the coherent group elements $$
| \vec n, j \rangle:=D^j(U_{\vec n})\, D^j(e)\,,
$$ 
with $e$ unit element of the group, $\vec n$ direction on $S^3$ that generically individuates $U\in SU(2)$ and $ D^j(e)\equiv |j, \pm \hat{z}j \rangle$. \\

This step allows to recover the DNN structure as the semiclassical limit of TQNN. Output 1-complexes (quantum spin-networks) and 2-complexes functorial structures in order to match the classical DNN structures must be evaluated on boundary coherent group elements. Furthermore, by recognizing that (\ref{funbis2}) retains an heat kernel for the SU$(2)$ group elements, the coherent group elements can be used as a basis for the functorial structure that defines the formula 
$$\mathcal{Z}_\mathcal{C}(U_l)=\int_{{\rm SU}(2)^{2(E-L)-V} } dU_{v(e)} \, \int_{{\rm SU}(2)^{\mathcal{V}-L}} dU_f\, \prod_f\, \mathcal{K}_{f*}(U_{e*},U_f)\,.
$$

The same must happen for (integrated) bulk coherent group elements. The structure of TQNN naturally encodes topological charges through the functorial quantum dynamics ensured by the 2-complexes, which create either vertices and then novel functions of intertwiner quantum numbers, or other topological charges encoded in the knotting and linking of the edges in the bulk of the 2-complex.\\

Specifically, we assume that the size of the training data is sufficient to select or, better, to learn specific paths in the boundary graph and bulk 2-complex within the most general available TQNN architecture. These paths are characterized by three different types of associated non-perturbative topological charges. These latter in turn provide the sub-structures that are involved in the generalization process, as a subset supported on general 2-complexes. The topological charges that are switched on over the learning process, together with the corresponding metric properties, implement effectively the generalization process. 
In this sense, our approach is expected to provide a solution to the problem as raised by Zhang et al, 2016. In particular: 

\begin{itemize}

    \item 
    The randomization of the labels of a TQNN state will not induce overfitting, as a consequence of the encoding of information achieved by the QNN through the topological invariants. The quantum nature of the QNN will induce fluctuations around values of the parameters to be estimated, in a way that is compatible with the zero assumption for these parameters. This assumption would instead change the topology of the graph, and thus affect the encoding of information by the QNN. As a consequence, the disappearance of topological features of the graphs will avoid the memorization by brute force of the training samples.

    \item 

    However, a DNN architecture will be trapped into an overfitting regime till memorizing the training examples by brute force, since by definition of DNN the training error vanishes --- the variance for the $j$ scale as $1/\sqrt{\bar{j}}$. In other words, corresponding DNN to a set of spin-network evaluated into coherent group elements, the associated training error is zero.
    
\end{itemize}

Contributions to the topological invariants can be recognized to be of several different types, including the ones associated to the connectivity of the graphs, the linking and the knotting (e.g. in the loops decomposition of the TQNN boundary and intermediate spin-network states) and the states' sum invariants. The first two classes will be local in the experimental implementation of the TQNN, while the latter represents a global charge, the analytical expansion of which in the deformation parameter might entail an infinite numbers of momentum expansion of the charge.
\\
\\
Notice that generic boundary states are characterized by two classes of parameters, which we dub as topological and metric parameters: As reminded above, the former ones are captured either by the topology of the graph, or by the topological invariant (linking and knotting) quantum numbers, which can be expressed in terms of quantum group representations and are characterized by the deformation parameter of the quantum group, while the latter ones are captured by the spin/label of the representation itself. Whenever not enough information about the topology is specified by the training data, any TQNN 2-complex with enough topological internal structure to account for the classification task will be selected. 
In other words, if the training data prescribe an effective shrinking of the ``measure'' of edges and links to zero, any topological feature of the graph, such as the valency of a node, or the knotting or linking of an edge, will cease to be. 
Metric parameters instead are individuated by the Gaussian weights associated to the coherent group elements assigned to the TQNN states, and recovered by fit on the spin representation set that is assigned to each training state. In this sense, since the parameters fit is achieved considering the whole amplitude $\mathcal{A}$, the resulting topology qualifies as a derivative-free feedforward architecture in which a composition of intermediate evolution operators among the hidden layers does not need to backpropagate the information.

\section{A new working hypothesis}

As a consequence of the previous discussions, we propose as working hypothesis for this proposal that the learning process of DNN shall be interpreted within an extended framework, which follows the very same axioms of quantum mechanics and quantum topology, through the formulation of TQFT. In other words, we see a TQNN as a quantization of a DNN whose $\hbar \rightarrow 0$ limit recovers the classical case. In the learning process of a TQNN, the substantial feature that a TQNN learns, is the selection of relevant geometries in the partition function that determines transition amplitudes utilized to classify. The main idea that constitutes the backbone of the present framework is that DNN should be addressed at the TQNN level. Training examples or tests samples will be captured by the spin representations of the TQNN quantum state, which are superpositions of the boundary Hilbert space elements. Moreover, we point out that TQNN implicitly carry a quantum computation perspective, since the boundary states in general are mixed as linear combinations of pure spin-network basis elements. Transition amplitudes will return the probability of a state as being in a certain spin-network basis state. The generic boundary states are characterized by two classes of parameters, which we dub topological and metric parameters: The former ones are captured by the topology of the graph, hence by the topological invariant (linking and knotting) quantum numbers, while the latter ones are captured by the spin of the representation itself. Pertaining to the topological parameters, information provided by the training samples, together with the definition of training error in terms of the internal product of boundary quantum states, substantially determines the structure of the bulk, and therefore the functor that determines transition amplitudes, in the learning process. We argue that the topological parameters are enough to learn the classifier, namely the TQNN 2-complex that provides the functorial structure of the TQNN, playing a similar role to the frequency threshold in the photoelectric effect: Whenever not enough information about the topology is specified by the training data, any TQNN 2-complex with enough topological internal structure will be selected. This might be considered as a TQNN counterpart of a similar phenomenon in the theory of TQFT, and its relations to Chern-Simons theory and the Jones polynomial. In fact, celebrated results of Witten \cite{Wit} has shown that the partition function associated to the action corresponding to Chern-Simons theory is independent of the metric, although the action itself is not. We have incurred in a similar situation, and we argue that the notion of generalization in TQNN theory and, as a limit, in DNN theory, lies precisely here. Although the partition functions that are used to determine the transition amplitudes are topological (hence the name TQNN), what is learnt during the learning process is what geometries to associate to given classified patterns. Metric parameters are individuated by the Gaussian weights associated to the coherent group elements assigned to the TQNN states. The size of the training set then will represent the analog of the intensity of the electromagnetic field in the photoelectric effect, namely the number of photons impinging the plates of the condenser: If the size of the training set is not sufficient, i.e. it does not include enough group elements, or the training set is too noisy, links and nodes will not be sufficient to learn any classifier. Lastly, the “richness” or “energy” of the set of labels allows to “switch on” the links, and thus the nodes and the topological linking and knotting invariants, only for non-trivial (non-zero) values of the spin. 

\section{Conclusions}
\noindent
Moving from the perspective of TQFT, we have defined the concept of ``Topological Quantum Neural Network'' and shown that that classical DNN can be seen as a subcase of TQNN, and emerge in a coherent group theoretical sense as a limit of TQNN. This allowed us to establish a dictionary translating a number of ML key-concepts in the terminology of TQFT. More importantly, we have proposed a framework that provides a working hypothesis for understanding the generalization behavior of DNN.

The novelty of our approach, particularly when compared to recent studies in the literature (\cite{5},  \cite{beer}), stands in taking into account fully, for the first time, the truly topological structure of graphs and 2-complexes on which the TQNN states are supported. Indeed, ours is not only a pictorial representation, in terms of graphs, of product states belonging to the total Hilbert space (Fock space) of the theory. Instead, what we have developed is a scheme that allows to associate ML concepts to topologically invariant features of the graphs (inter-connectivity of edges, linking and knotting numbers, topological invariants on 2-complexes) and 2-complexes involved in the TQNN construction.

A number of further lines of research could be pursued starting from our approach:
\begin{itemize}
    \item 
    [1.] Providing empirical results concerning the working hypothesis previously described so to corroborate the claim that the notion of generalization introduced in this article is consistent;
\item [2.] Defining new complexity measures more appropriate to the framework we described and adequate to explain the behavior of over-parametrized models such as DNN. It would also be of interest to pursue deeper experimentation with variety of benchmark data sets, so to relate complexity measures to concrete examples;
\item [3.] Introducing the notion of ``time'' into the architecture by modelling phenomena of the cortical plasticity such as firing rate or spike timing, see \cite{sjostrom}. In particular, this perspective implies the necessity of using TQFT that have one extra dimension with respect to the concrete ones that have been used in this article. The basic theory does not change, in that the notion of TQNN does not require fixing a specific dimension in the cobordism category, but the corresponding algebro/analytical machinery certainly becomes heavier.
    
\end{itemize}
 
\section*{Acknowledgements}
AM acknowledges support by the NSFC, through the grant No. 11875113, the Shanghai Municipality, through the grant No. KBH1512299, and by Fudan University, through the grant No. JJH1512105. NG acknowledges Foundation of the Jiangsu Higher Education Institutions of China Programme Grant 19KJB140018 and XJTLU REF-18-02-03 Grant. ML acknowledges the support from National Science Foundation of China Grant No.~12050410244. EZ was supported by the Estonian Research Council through the grant MOBJD679.
\appendix

\section{Topological Quantum Field Theory}

We provide in this appendix a deeper introduction to Topological Quantum Field Theory (TQFT), spin-network (boundary) states and (bulk) 2-complexes functorial evolution of boundary states.       

\subsection{Classical phase-space and spin-network states}

The theory is the principal SU(2)-bundle over a \emph{D}-dimensional base manifold $\mathcal{M}$. The SU$(2)$-connection $A$ realizes the parallel transport among infinitesimally closed fibers of the principal bundle. The parallel transport along a finite path $\gamma$ connecting any two points of $\mathcal{M}$ is individuated by 
\begin{equation}
    H_\gamma[A]= {\rm P}\, e^{\int_\gamma A}\,,
\end{equation}
which denotes the path ordered exponential $P$ of the integrated flux of $A$ along $\gamma$. The holonomy then provides a group element $g\in {\rm SU}(2)$. The trace of the holonomy along a closed path (a loop $\alpha$) can be expanded, taking into account a squared loop of infinitesimal edge $\epsilon$, as 
\begin{equation}
  \lim \limits_{|| \alpha || \rightarrow 0}  W_\alpha[A] =  1\!\!1 - \epsilon^2 F[A] +\dots\,,
\end{equation}
where $||\alpha||$ denotes the measure of the loop $\alpha$, and   $F[A]=dA+A\wedge A$ is the field strength, or curvature, of the connection $A$. The connection $A$ is both a \emph{1}-form on $\mathcal{M}$ --- indeed, its curvature is a \emph{2}-form over $\mathcal{M}$, since the differential $d$ is one-form --- and an element of the $\mathfrak{su}(2)$ algebra. Thus, it admits the decomposition over the generators $\tau^a$, with $a=1,2,3$ indices in the adjoint representation of the algebra. Consequently, the connection $A$ and its curvature $F[A]$ acquire the dependence on the internal indices, respectively $A=A^a \tau^a$ and $F^a[A]=d A^a + \epsilon^{abc} A^b\wedge A^c$, the Levi-Civita symbol $\epsilon^{abc}$ providing the structure constants of SU$(2)$ and the Einstein convention of summing repeated indices is intended. \\ 
\\
A TQFT can be introduced considering the topological action associated to the Lagrangian density function  
\begin{equation} \label{act}
  \mathcal{L}[A] = B^a \wedge F^a[A] = \rm{Tr}[B \wedge F[A]]\,,
\end{equation}
where the $B$ field denotes a $\mathfrak{su}(2)$ algebra valued \emph{D}-form, which is the canonically conjugated momentum to the connection $A$, and the trace over the generators of the algebra is normalized to the identity and yields $\rm{Tr}[\tau^a \tau^b]=\delta^{ab}$. The phase-space variables $A$ and $B$ can be then paired in a symplectic construction, imposing the Poisson brackets
\begin{equation}
\{A^i_a(x_1), B^b_j(x_2) \}= \delta_a^b  \, \delta^i_j  \, \delta(x_1,x_2)\,,
\end{equation}
with $i=1,\dots, \emph{D}$ space indices over the dimensions of $\mathcal{M}$.\\
Holonomies realize the smearing of the configuration space variables, i.e. the connections $A$, along the paths $\gamma$. \\
\\
Similarly, the smearing of the frame fields $B$ can be implemented by substituting their fluxes calculated through the surfaces $\Sigma$ of co-dimension $1$ that crosses the paths $\gamma$ at least in one point, namely 
\begin{equation}
B_\Sigma=\int_\Sigma B\cdot n \,,
\end{equation}
where $n$ is the normal to the surface $\Sigma$ and the dot denotes contraction of indices. For example, since the dimension of the path $\gamma$ is $1$, its co-dimension $1$ surface in a \emph{3D} ambient space will be a \emph{2D} surface. 

The theory we just introduced retains what is called a gauge symmetry, namely a symmetry under internal transformations, which individuates an equivalence class that describes an observer. These are instantiated by transformations involving generic group elements $g\in {\rm SU}(2)$, i.e. 
\begin{equation} \label{g1}
A \rightarrow A_g = g^{-1} A g + g^{-1} d g\,, 
\end{equation}
and 
\begin{equation} \label{g2}
B \rightarrow B_g = g^{-1} B g\,.
\end{equation}
It is trivial to check that the action (\ref{act}) is invariant under the joined action of (\ref{g1})-(\ref{g2}). The infinitesimal expansion of finite transformation rules (\ref{g1})-(\ref{g2}) can be cast at the $\mathfrak{su}(2)$ algebraic level, through the infinitesimal expansion of a group element around the identity, i.e. $g\simeq 1\!\!1 + \alpha^a \tau^a + \dots$. This individuates an infinitesimal transformation 
\begin{equation} \label{g3}
\delta_\alpha B = [B,\alpha]\,, \qquad \delta_\alpha A = \mathcal{D}_A \alpha\,,
\end{equation}
where the commutators $[\,,\,]$ denote the adjoint action of the algebra. The generators of the algebra appear in $ B=B^a \tau^a $ and $\alpha=\alpha^a \tau^a$, while $\mathcal{D}_A$ denotes the covariant SU$(2)$ derivative $\mathcal{D}_A:=d + A$. \\
\\
Another symmetry, which is relevant for the definition of TQFT, is the shift symmetry. This is actually ensuring the theory under consideration to be topological, as it is straightforward to recognize by looking at 
\begin{equation} \label{t1}
B\rightarrow B+\delta_\eta B \,, \qquad  \delta_\eta B= \mathcal{D}_A \eta \,,
\end{equation}
and
\begin{equation} \label{t2}
A\rightarrow A+\delta_\eta A \,, \qquad  \delta_\eta A=0\,,
\end{equation}
where $\eta$ is any arbitrary infinitesimal $0$-form (a function). Under the infinitesimal transformations (\ref{t1})-(\ref{t2}), the variation of the action of the theory $\mathcal{S}[A]=\int_\mathcal{M} \mathcal{L}[A]$, namely
\begin{eqnarray}
\delta_\eta \mathcal{S}[A,B]=\mathcal{S}[A+\delta_\eta A, B+\delta_\eta B]-\mathcal{S}[A,B]\,,
\end{eqnarray}
vanishes, due to the Bianchi identity $\mathcal{D}_A F[A]=0$. This latter identity appears in the variation of the action due to an integration by part:
\begin{eqnarray} \label{t3}
\int_\mathcal{M} {\rm Tr} [(B + \delta_\eta B ) \wedge F[A+ \delta_\eta  A] &=& \int_\mathcal{M} {\rm Tr} [(B + \mathcal{D}_A \eta ) \wedge F[A] = \nonumber \\
 \int_\mathcal{M} {\rm Tr} [B \wedge F[A]] - \int_\mathcal{M} {\rm Tr} [B + \wedge  \eta \mathcal{D}_A  F[A] &=&  \int_\mathcal{M} {\rm Tr} [B \wedge F[A]] \,.
\end{eqnarray}
This symmetry is often referred to as a ``gauge symmetry'' of the $BF$ theory, which individuates a class of equivalence among physical solutions that differ by this transformation. 

On the other hand, the equation of motions are specified by the variation of the action with respect to the phase-space fields: \begin{eqnarray} \label{t11}
\mathcal{D}_A B=0\,, \qquad \qquad F[A]=0  \,.
\end{eqnarray}
Solutions are then "flat", or with zero curvature, i.e. $F[A]=0$, while the frame fields satisfy the Gau\ss\, constraint $\mathcal{D}_A B=0$, which generates the gauge transformations. Locally, by the topological shift symmetry, any frame field $B$ that satisfies the Gau\ss\, constraint can be recast as $\mathcal{D}_A  \eta$, for some $\eta$. This is true as locally closed forms are exact, and continue to satisfy the Gau\ss \,  constraint. This implies that locally the solutions of the equations of motion belong to the same equivalence class, modulo gauge transformations and shift symmetry transformations. Since these can be mapped into vanishing configurations, this argument finally shows that there are no propagating degrees of freedom in $BF$ theories, namely that these theories are topological. 

\subsection{Graph-kinematics} \label{graki}
\noindent 
As a last step before proceeding to the definition of the 1- and 2-complexes, we introduce the irreducible representations of the group, the so-called “spin” numbers, and the inter-twiner numbers, depending on the SU$(2)$ recoupling theory. At this purpose, we remind that in this case holonomies over a path $\gamma$ are group elements of SU$(2)$, and thus undergo the transformations 
\begin{equation} \label{hg}
H_\gamma[A] \rightarrow g^{-1}_{s(\gamma)} H_\gamma[A] g_{t(\gamma)}
\,,
\end{equation}
where $g_{s(\gamma)}$ and $g_{t(\gamma)}$ are group elements assigned respectively to the source and the target of an oriented path $\gamma$. For SU$(2)$, irreducible representation of holonomies are provided by the Wigner matrices and labelled by the semi-integer $j$-spin numbers, namely  
\begin{equation} \label{t21}
D^{(j_\gamma)}(U_\gamma) \,, \qquad U_\gamma \equiv H_\gamma[A]
\,,
\end{equation}
SU$(2)$ intertwiners are expressed as the (group elements) integrals of a number of copies of irreducible representations (Wigner matrices). As a compact group, SU$(2)$ is endowed with a Haar measure (invariant under gauge transformations and coordinate reparametrizations) that enables the definitions of the intertwiner invariant tensors. These latter quantities can be thought to be associated to the nodes where endpoints (target points) and origins (source points) of the paths $\gamma$ intersect. A collection of $n$ path $\gamma_1, \gamma_2 \dots \gamma_n$ intersecting at their target and source points (nodes) provides a graph $\Gamma$. The internal indices of the Wigner matrices integrated ensure gauge-invariance through the contraction with the holonomies flowing across the node. Integrating in the Haar measure the irreducible representations of the holonomies, the target or source points of which cross at the node, and which are labelled by the spin $j_{\gamma_1}, j_{\gamma_2}\dots j_{\gamma_n}$, provided the expression for the inter-twiner 
\begin{equation} \label{int}
v_\iota = \int_{\rm SU(2)} dU\,  D^{(j_{\gamma_1})}(U) \, D^{(j_{\gamma_2})}(U) \dots D^{(j_{\gamma_n})}(U)\,,
\end{equation}
having again suppressed all the (intertwiner and Wigner matrices) representation indices.\\
\\
A collection of holonomies, the internal indices of which are contracted with the intertwiners defined by integration of the group elements at the nodes, defines a spin-network state. In terms of its constituents, the holonomies and the intertwiners, a spin-network state cast as
\begin{equation} \label{sn}
\psi_{\Gamma,\{j_l\}, \{\iota_n\} }[A]= \left( \bigotimes_{n \in \Gamma}  v_{\iota_n} \right) \cdot \left( \bigotimes_{{\gamma_l} \in \Gamma}\, \stackrel{j_{{\gamma_l}}}{D}(U_{\gamma_l} [A]) \right) \,,
\end{equation}
where the dot denotes the contraction of internal indices, and $l=1,\dots n$ label the $n$ paths $\gamma$ that compose the graph $\Gamma$. 

\begin{figure}
\centering
\includegraphics[width=10 cm]{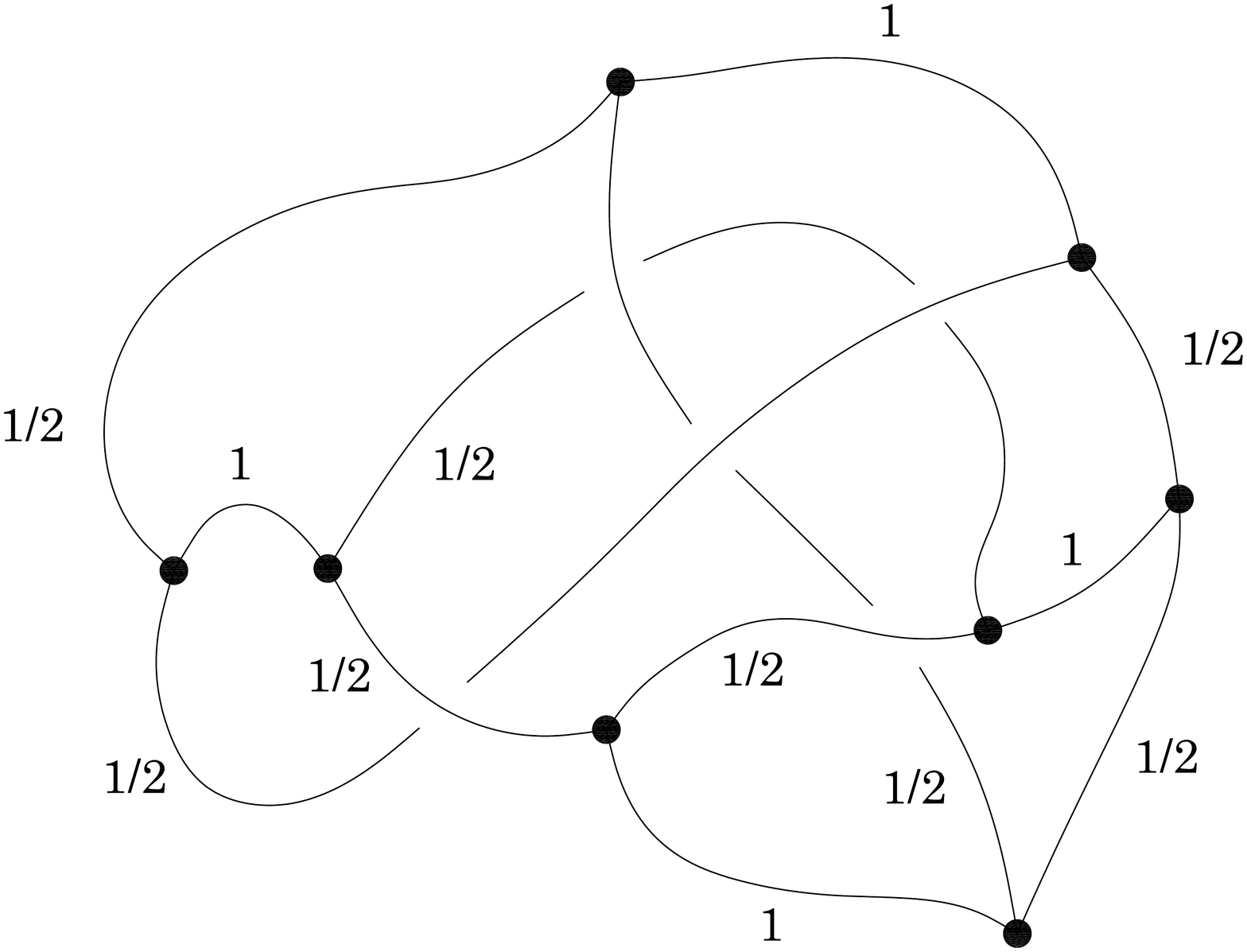}
\caption{A graph with tri-valent nodes colored under SU$(2)$.}
\end{figure}

SU$(2)$ spin-network states are equipped with a Haar measure, which ensures invariance under gauge transformations and diffeomorphisms (coordinate reparametrizations) on the base manifold $\mathcal{M}$, of the internal product
\begin{equation} \label{inpro}
<\Psi_{\Gamma', j_{\gamma}', \iota_n'}[A]|\, \Psi_{\Gamma, j_{\gamma}, \iota_n}[A] >= \delta_{\{ \Gamma ' \}, \{ \Gamma \} } \delta_{j_\gamma ', j_\gamma  } \delta_{\iota_n ', \iota_n  }
\end{equation}
Invariance under diffeomorphisms, which is expressed by the Kronecker delta between classes of equivalence of graphs endowed with the same topology, namely $\{\Gamma\}$, instantiates the symmetry under elastic transformations, rendering the graph structure truly topological. In this study, graphs $\Gamma$ are also referred to as 1-complexes.\\
\subsection{Graph-dynamics} \label{grady}
\noindent 
A concept of dynamics requires the definition of boundary states (1-complexes), the quantum evolution of which is provided by relative transition amplitudes. These are captured by the path integral (realizing the vacuum-vacuum transition, with no underlying graph structure) and the expectation values in its measure.
It is convenient to introduce the mathematical concept of 2-complex $\mathcal{C}$. A 2-complex $\mathcal{C}$ is composed by edges $e$ departing or ending either at nodes $n\in \Gamma$ or at vertices $v$ internal to $\mathcal{C}$, by faces $f$ bounded by either links $\gamma$ or internal edges $e$, and vertices $v$ where edges cross. We are going to show how to associate a functor --- either the partition function $Z_\mathcal{C}[U_{\gamma_l}]$, or the expectation value of boundary state in the path-integral associated to the topological theory --- to a 2-complex $\mathcal{C}$ endowed with boundary group elements $U_{\gamma_l}$. 

The partition function for the BF model over a SU$(2)$-bundle is specified by the expression
\begin{equation}
\mathcal{Z}=\int \mathcal{D}A \mathcal{B} \, e^{\imath \int_\mathcal{M} {\rm Tr}[B\wedge F]} =    \int \mathcal{D}A ``\delta(F)''.
\end{equation}
where in the last equality we introduced a Dirac delta measure on the space of flat connections. This is understood \cite{Baez_2000}
from smearing the phase-space variables and then casting the partition function as 
\begin{equation} \label{pf}
\mathcal{Z}(\Delta)=\int_{{\mathfrak{su}(2)}^E} \prod_{e\in E} dB_e \, \int_{{{\rm SU} (2)}^{E^*} } \prod_{e\in E^*} dU_e\, e^{\imath 
\sum_{e\in\Delta} {\rm Tr[B_e F_e]}}\,,
\end{equation}
where $\Delta$ denotes the triangulation of the manifold $\mathcal{M}$ --- this allows to introduce a simplicial complex $\Delta^*$ that is dual to the triangulation $\Delta$ --- $E$ denotes the set of edges $e$ of the triangulation $\Delta$, and $E^*$ the set of edges $e*$ of the dual simplicial complex $\Delta^*$. Furthermore, in the expression (\ref{pf}) we have been using the natural definition of the curvature, which is expressed by the product of group elements $U_{e*}$ associated to the links around the boundary $\partial f*$ of a dual face $f*$ (thus associated with the dual face itself):
\begin{equation}
U_{f*}=\prod_{e^*\in \partial f*}U_{e*}\,.    
\end{equation}
where $F_e=\ln U_{f^*}$, namely individuates a Lie algebra element that entails the discretization of the connection field curvature on the edges $e$ of $\Delta$.
Integration over the algebra elements $B_e$ provides the expression for the Dirac delta on the product of group elements that realizes the smearing of the curvature, namely    
\begin{equation}
\int_{{\mathfrak{su}(2)}^E} \prod_{e\in E} B_e \, e^{\imath \sum_{e\in\Delta} {\rm Tr[B_e F_e]}} = \delta(e^{F_e})=\delta(U_{f*})\,.
\end{equation}
The partition function then casts
\begin{equation}
\mathcal{Z}(\Delta)= \int_{{{\rm SU} (2)}^{E^*} } \prod_{e\in E^*} \, dU_{e*} \prod_{f*} \delta(U_{f*})\,.
\end{equation}
This formula  finally admits a re-manipulation in terms of the irreducible representation of SU$(2)$, which thanks to the Peter-Weyl expansion, is provided by Plancherel formula  
\begin{equation}
 \delta(U_{f*})=\sum_{j_{f*}} \Delta_{j_{f*}} \chi^{j_{f*}}(U_{f*}) \,,
\end{equation}
where $j_{f*}$ denote half-integer numbers that label SU$(2)$ irreducible representations, $\Delta_j=(2j+1)$  the dimension of these latter, and $\chi^{j}(U)= D^{j}(U)^\alpha_\alpha$ is the character of the group element $U\in SU$(2), i.e. the trace of a Wigner matrix over the internal indices $\alpha$ in the representation Hilbert space. Then the partition function recasts 
\begin{equation}
 \mathcal{Z}(\Delta)= \sum_{j_{f*}} \int_{{{\rm SU} (2)}^{E^*} } \prod_{e\in E^*} dU_e\, \prod_{f*} {\rm Tr}[D(\prod_{e^*\in \partial f*}U_{e*})]\,,
\end{equation}
which depends only on the recoupling theory of SU$(2)$, and retains a dependence on the dimension of the manifold $\mathcal{M}$, in which both the graphs $\Gamma$ and the 2-complex $\mathcal{C}$ are merged. Thus, we can identify the no-boundary path-integral amplitude $\mathcal{Z}(\Delta)$ with the no-boundary functor $\mathcal{Z}_\mathcal{C}$, i.e.
\begin{equation}
\mathcal{Z}_\mathcal{C}=\mathcal{Z}(\Delta)\,
\end{equation}
where there is no dependence on the boundary group elements. 

\bibliography{main.bib}
\end{document}